\documentclass[a4paper,12pt]{article}

\usepackage[utf8]{inputenc}
\usepackage{graphicx}
\usepackage{subfig}
\usepackage{fullpage}
\usepackage{hyperref} 

\title{Electronic transport through molecular transistors in the polaronic regime}
\author{Rubén Seoane Souto}

\author{R. Seoane Souto\thanks{\href{mailto:ruben.seoane@uam.es}{ruben.seoane@uam.es}}, A. Levy Yeyati, A. Martín-Rodero and R. C. Monreal\\
\small Departmento de Física Teórica de la Materia Condensada,\\
\small Centro de Investigación de Física de la Materia Condensada (IFiMaC)\\
\small and Instituto Nicolás Cabrera\\
\small Universidad Autónoma de Madrid, E-28049 Madrid, Spain\\[-0.8ex]
\date{}
}

\begin{document}
\maketitle

\begin{abstract}
In this work, a new theoretical approach to study the non-equilibrium transport properties of nanoscale systems coupled to metallic electrodes with strong electron-phonon interactions is presented.
The proposed approach consists in a resummation of the dominant Feynman diagrams from the exact preturbative expansion. This scheme is compared with methods that can be 
found in the literature. It shows a good agreement with these methods in the range where they are known to provide good results, for a wide range of parameters. Also, it is compared with path-integral Monte Carlo
 calculations giving a relatively good agreement for polaronic and non polaronic regimes. Some preliminary results for the current noise obtained with our method are presented.
\end{abstract}

\section{Introduction:}
In the past decades, there has been an increasing interest in the fabrication of smaller and faster electronic devices. Until now, electronics has been mainly developed using 
inorganic materials, where silicon has played a fundamental role. An example of this interest is the so called Moore's law, which shows that the maximum number of transistors in a chip grows 
exponentially with the time. However, limitations shown by this technology have converted molecular electronics in a very interesting field for applications.\newline

The small dimensions of molecular circuits together with the great variety of electrical, mechanical and optical properties, make this field attractive for basic research.
These molecules are ideal systems where fundamental electron transfer mechanisms, which play a fundamental role in chemistry and biology, can be studied. Understanding 
these mechanisms could lead to new functionalities that are impossible to implement with conventional devices. The molecules used can go from the smaller ones (few nanometers in
size) to big polymeric ones.\newline

Historically, the first study of charge transferred between molecules in the 40s done by R. Mulliken and A. Szent-Gyorgi, is considered to be the birth of molecular electronics.
 Nevertheless, the first proposal for practical applications was made in 1974 by Arieh Aviram and Mark A. Ratner \cite{primera}. In that work they proposed the fabrication of 
a current rectifier using an organic molecule. At the end of the 1980's and the beginning of the 1990's, the appearance of the metallic atomic-sized contacts had an 
important impact in this field. Since then, a huge theoretical and experimental effort has been done on understanding these kind of systems \cite{JCC}.\newline

The reduced size of these systems could lead to another interesting advantages, such as lower cost and higher efficiency. Most of the molecules have a poor conductance and could
reduce the operation time. Moreover, the huge amount of degrees of freedom of a molecule (charge, rotation, vibration, conformational,..) could lead to new electronics. In addition, the conductance
 of the molecule depends strongly on its interactions with the electrodes and also on the vibrational degrees of freedom. So, the study of the properties of such a system (a molecule between 
two leads) could provide a method to characterize molecules. In reference \cite{ADN} there is an example of these applications, where the authors propose a design for a DNA reader.\newline

\begin{figure}
  \centering
	\includegraphics[width=1\textwidth]{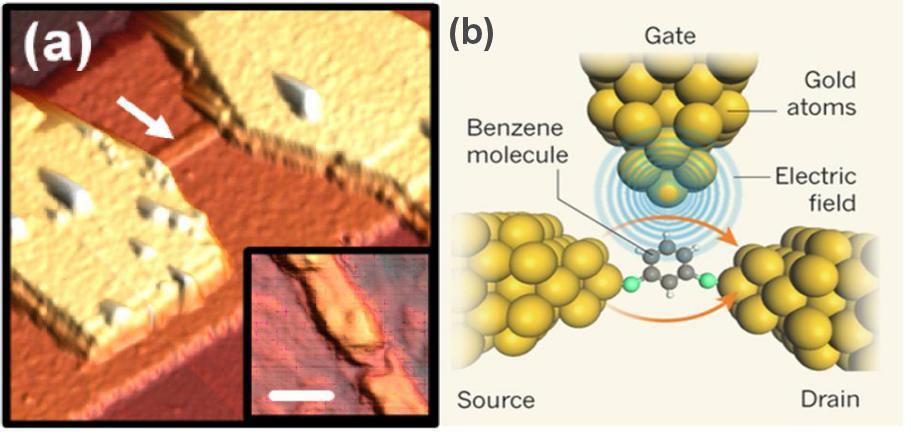}
  \caption{\small{\emph{A.} microscopic image of the gap where molecule is located. \emph{B.} Scheme of a typical setup for molecular electronics \cite{rotura}.}}
  \label{molecula1}
\end{figure}

The usual experimental setup for studying electronic transport through molecules is similar to the one of a traditional transistor (figure \ref{molecula1} b). Transport phenomena takes place between
the two leads, with a nanometer gap where the molecule is located (figure \ref{molecula1} a). A third electrode can be used as a \emph{gate}, to tune the displacement of molecular energy 
levels with respect to Fermi level of the electrodes.\newline

The aim of this work is the theoretical study of the electronic transport through a nanoscaled system coupled to a vibrational degree of freedom. The molecule has a possibility to vibrate between the two 
leads, because of Van der Waals interaction with them. Effects of this kind of vibration, which has very low energies compared with the coupling to the electrodes, have been extensively studied
\cite{shuttling}. However, the situation that will be studied in this work is a different one, and considers the coupling to an internal vibration, which has typically higher energies ($\hbar\;\omega_0\;\simeq$ 100 meV). 
This internal vibrations have demonstrated to be important for quantum dots made by carbon nanotubes \cite{carbono,carbono2}, and also for some systems fabricated by mechanical break
junctions \cite{rotura}.\newline

During the last years, many groups have studied this system in different regimes and using different approximations. One important parameter that defines the regime is the
 \emph{tunneling rate} ($\Gamma$), which measures the average lifetime of an electronic state in the molecule due to the coupling with the electrodes. The case that is going to be 
studied is when this coupling is much higher than thermal excitations ($\Gamma\gg k_b T$). This regime, known as \emph{quantum regime}, the coherence is maintained during the transport
of electrons.\newline

This internal vibration of the molecule produces a coupling with electrons, measured by a parameter $\lambda$, that bring new processes and a very rich physics. 
As a function of this parameter, two regimes can be distinguished. A first one, where $\Gamma\gg\omega_0\gg\lambda^2/\omega_0$, has been extensively studied using atomic wires 
\cite{atomic_wire,shot1} and single molecules\cite{rotura}. In this regime a perturbative approach to the lowest order in $\lambda$ is appropriate \cite{Laura,review,Mitra,Egger,Wohlman,Remi}. 
In general, the effect of the internal vibration in the transport properties of the system in this regime is a jump of about 1\% in the conductance when the mode is excited.\newline

 The regime analyzed in this thesis is the opposite one ($\Gamma\ll\lambda^2/\omega_0$), where the coupling of the electrons with the vibration is strong. This regime is called \emph{polaronic
regime}. This problem has been studied using some approximations that are valid only for a given range of parameters\cite{Galperin,soloRemi,paperPTA,Flensberg}. 
The motivation of this work is to describe the behavior of the system in a 
range of parameters as wide as possible. In section \textbf{2} the basic mathematical tools will be introduced and some existing approximations described; in section \textbf{3} the new scheme is 
proposed; in section \textbf{4} we analyze the results of the approximation and compare them with other methods and quantum Monte Carlo calculations; and, in section \textbf{5}, we present the 
concluding remarks.

\section{Theoretical methods:}

The work developed is focused on the calculation and prediction of macroscopic observables, like the current, using the knowledge about the microscopical world. A huge amount and variety
of interactions between particles lead to some emerging phenomena. The theoretical tools used for the study of this systems are \emph{many body} techniques \cite{Mahan}. The formalism used is the non
equilibrium Green's functions (Keldysh formalism). These Green's functions are useful because every average value of any observable can be written in terms of them. At zero temperature, the retarded 
component of Green's function can be expressed as:
\begin{equation}
  \label{Gretarded}
  G^R(x,t;x',t')=-\frac{i}{\hbar}\theta(t-t')\left\langle\Psi_0\left|\left[\hat{\psi}(x,t)\hat{\psi}^{\dagger}(x,'t')\right]_+\right|\Psi_0\right\rangle
\end{equation}
where  $\left|\Psi_0\right\rangle$ represent the ground state of the system described with $H$, time-independent Hamiltonian. $\hat{\psi}(x,t)$ and $\hat{\psi}^{\dagger}(x,'t')$ are the electronic field operators.
 With this definition, the Green's function, with $t>t'$, can be interpreted as the probability amplitude of a particle to be created at time $t'$ in \textbf{r'} position and $\sigma'$ spin, and to be 
annihilated at $t$ in \textbf{r} and with $\sigma$ spin.\newline

This Green's function can be also related with the Keldysh components
\begin{equation}
  \label{retarded-keldysh}
 G^{R}(x,t;x',t')=\theta(t-t')\left(G^{-+}(x,t;x',t')-G^{+-}(x,t;x',t')\right)
\end{equation}
where the indices $+$ and $-$ denote the Keldysh branch (causal or anticausal) where the time is.\newline

It is often useful computing Green's functions from perturbation expansions, using Wick's Theorem. In this formalism, collective Green's functions can be written in 
terms single particle ones in perturbation theory. This infinite sum, represented by a series of diagrams, can be simplified and reduced to compute the Green's function from the Dyson equation:
\begin{equation}
   \label{Dyson1}
 \hat{G}=\hat{G_0}+\hat{G_0}\;\hat{\Sigma}\;\hat{G}
\end{equation}
where $\hat{\Sigma}$ is the self-energy, that represents the sum of the contribution of every irreducible diagram. In this term all the many-body effects appear implicitly
and it depends on the definition given to the perturbation term in the Hamiltonian, $V$.\newline

Finally, observables of the system can be obtained from these Green's functions. A function that will be interesting for the study of the problem is the 
\emph{spectral function}, also called \emph{density of states}, that can be written in frequency domain (Fourier transform) as:
\begin{equation}
 \label{espectral}
 A(x,x';\omega)\equiv-\frac{1}{\pi}\;\mbox{Im}\left[G^{R}(x,x';\omega)\right]; \quad x=x'
\end{equation}

\subsection{Modeling:}
\begin{figure}
  \centering
	\includegraphics[width=1.0\textwidth]{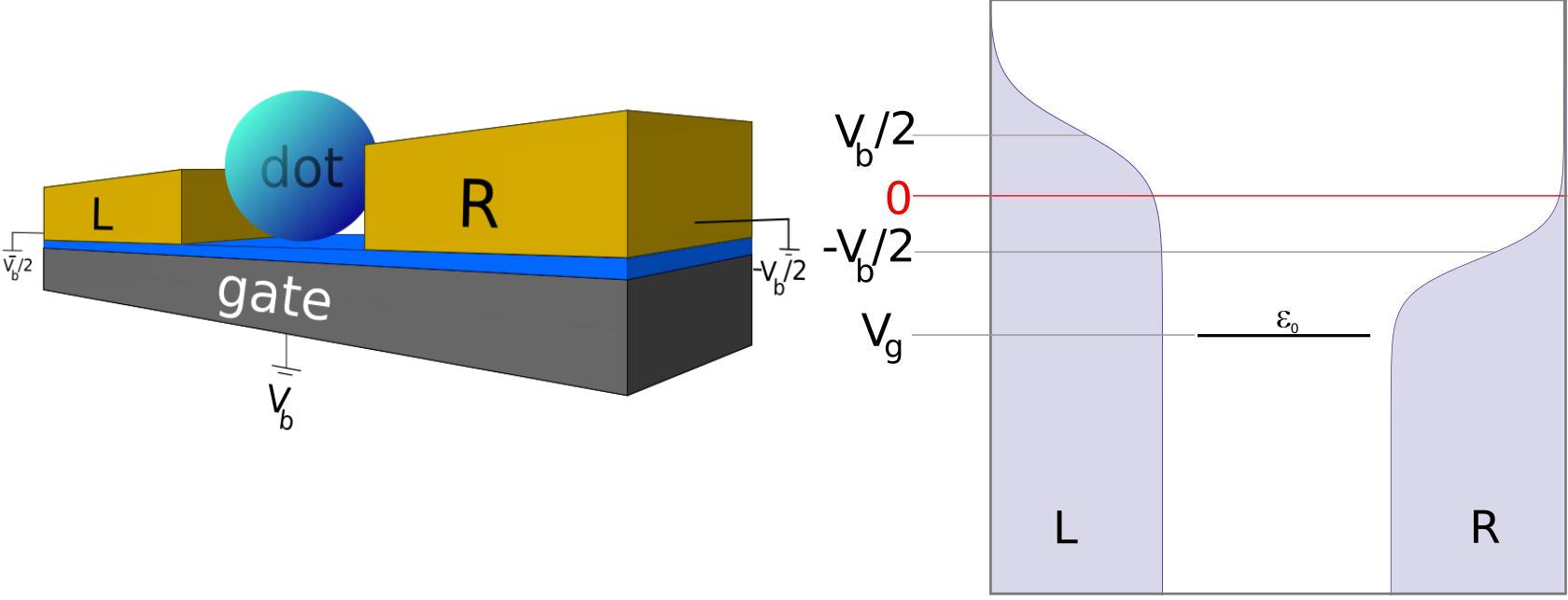}
  \caption{\small{\emph{A.} Scheme for a quantum dot system, with the voltages applied. 
\emph{B} Scheme of levels of energy of the minimum model. Shadowed regions represent the Fermi distributions of both leads (left and right) with an applied bias voltage of $V_d$. Continuous line
represents the single level of the quantum dot.}}
  \label{diagramaqd}
\end{figure}

The aim of this section is to define a model Hamiltonian to describe the system. The first problem that is going to be analyzed is that of a molecule between two electrodes, but without
the internal vibrational degree of freedom. This problem is the only one that can be solved exactly, and it will be useful to define some limits. A generalization of this kind of system is the 
quantum dot, that suppose small regions of matter placed in between electrodes that behaves as artificial atoms. A quantum dot is a nanometer device where electrons are confined and their energy 
levels are well separated (more than possible external perturbations). The minimum model that describes this kind of systems is given by next Hamiltonian 
(figure \ref{diagramaqd} a):
\begin{equation}
 H=H_L+H_R+H_{QD}+H_T
\end{equation}
where $H_T$ is the hopping term that allows electrons to jump from the electrodes to the quantum dot. $H_L$ and $H_R$ represent the terms of the Hamiltonian for both electrodes
(left and right), and $H_{QD}$ is the Hamiltonian of the isolated quantum dot.\newline

A very important property of the quantum dots is the so called \emph{Coulomb Blockade effect}. This effect is due to coulomb repulsion which leads a huge increasing if one more electron is added 
to the quantum dot (going from $N$ to $N+1$ electrons confined in the dot). In this regime the current across the system takes place electron by electron through last unoccupied energy 
level. For modeling the system, then, only one energy level will be considered (single level approximation). With this simplification, $H_{QD}$ and $H_T$ elements of the Hamiltonian can be written 
as:
  \begin{eqnarray}
    H_T&=&\sum_\textbf{k} t_L\left(c_{\textbf{k}_L}^{\dagger}\;d+d^\dagger\;c_{\textbf{k}_L}\right)+t_R\left(c_{\textbf{k}_R}^{\dagger}\;d+d^\dagger\;c_{\textbf{k}_R}\right)\nonumber\\
    H_{QD}&=&\epsilon_0\;d^\dagger\;d
  \end{eqnarray}
where $d$ is the annihilator operator of electrons in the quantum dot, and $c_{\textbf{k}_R}$ and $c_{\textbf{k}_L}$ the ones for the electrodes, for an electron with $\textbf{k}$.
$t_L$ and $t_R$ are the hopping elements corresponding to both electrodes and $\epsilon_0$ is the energy of the level of the molecule.\newline

\begin{figure}
  \centering
	\includegraphics[width=1.0\textwidth]{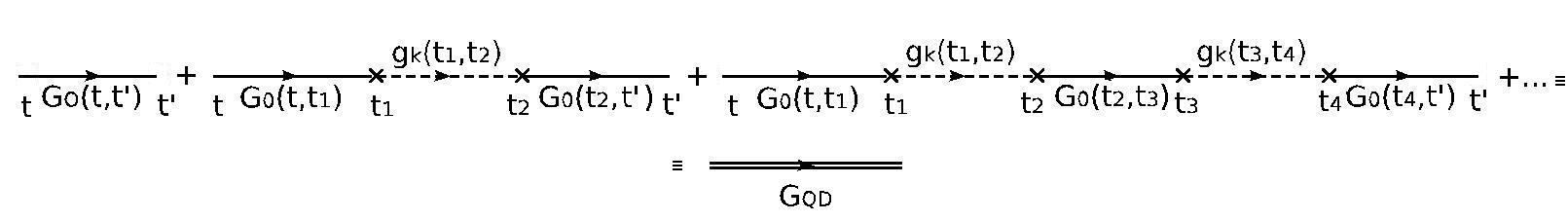}
  \caption{\small{Series of Feynman diagrams of a quantum dot with a single level, coupled to electrodes. $G_0$ is the Green's function of the isolated dot (\ref{g0}), 
and $g_k$ represents the Green's function of both electrodes. The crosses at the end of some of the lines represent hopping events.}}
  \label{QD}
\end{figure}

The position of the level can be controlled by applying a gate potential ($V_G$). Also, the position of the Fermi level of the electrodes can be controlled by applying a bias potential 
(figure \ref{diagramaqd} b) to the junction. The Green's function of the quantum dot system can be then computed exactly using Dyson equation \cite{Mahan}. Diagrammatically, this is equivalent to 
sum the series represented in figure \ref{QD}, up to infinite order.

The definition of the retarded Green's function in non equilibrium formalism:
\begin{equation}
 G^{R}_{QD}=-i\;\theta(t-t')\left\langle\left[d(t),d^\dagger(t')\right]_+\right\rangle
\end{equation}

Using this definition, we obtain the retarded Green's function:
\begin{equation}
	G_{QD}^R(\omega)=\frac{1}{\omega-\epsilon_0-\Sigma^{R}_0(\omega)}
\end{equation}

The main part that describes all the possible processes is the self-energy, that can be written as:
\begin{equation}
 \Sigma^{R}_0=\sum_{\textbf{k},\mu=L,R} V_{\textbf{k},\mu} \;g_{\textbf{k},\mu}^R \;V_{\textbf{k},\mu} 
\end{equation}

To simplify the calculation, the so called \emph{Wide Band Approximation} (WBA) is commonly used. This approximation consists on neglecting the real part of the self-energy.
This makes sense if the bands of the electrode are wide enough.
\begin{eqnarray}
 \label{auto}
 \Sigma^R&\approx&-i(\Gamma_L+\Gamma_R)\nonumber\\
 \Gamma_\mu&\equiv&\pi\;\sum_\textbf{k}\;V{^2}_{\mu,\textbf{k}}\rho_\mu(\epsilon_F);\qquad\mu=L,R
\end{eqnarray}
where $\Gamma_\mu$ is the \emph{tunneling rate} of the electrons between the molecule and the two leads. It is a measure of the average lifetime of the electronic states in the molecule.\newline
 
So, the retarded Green's function of the quantum dot can be finally written as follows:
\begin{equation}
	\label{g0}
	G_{QD}^R=\frac{1}{\omega-\epsilon_0-\Sigma^{R}_0}=\frac{1}{\omega-\epsilon_0+i\Gamma}
\end{equation}
where $\Gamma\equiv\Gamma_L+\Gamma_R$

\subsection{Anderson-Holstein model:}
\begin{figure}
  \centering
	\includegraphics[width=1.0\textwidth]{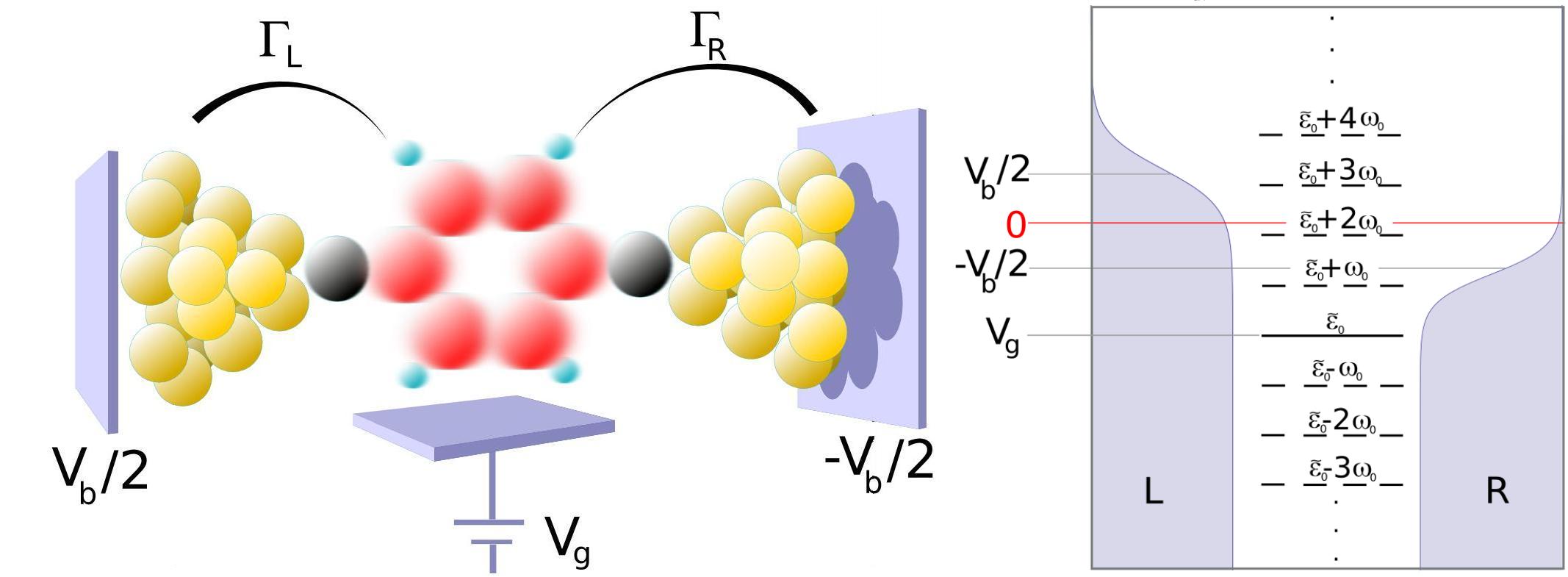}
  \caption{\small{\emph{A.} System scheme of a molecular quantum dot. \emph{B} Energy level scheme of the problem, once Lang-Firsov transformation of the Hamiltonian has been done (\ref{LF}). 
The level of the molecule appears repeated at integer values of the vibrational energy.}}
  \label{diagramaAH}
\end{figure}

Once the problem of a single level quantum dot has been introduced, the case of a molecule with a vibrational degree of freedom and placed between two electrodes is going to be analyzed
(figure \ref{diagramaAH} a). The minimum model that describes transport in this problem is the spinless Anderson-Holstein model, described by:
\begin{equation}
	H_{AH}=\omega_0\;a^\dagger\;a+\lambda\;d^\dagger\;d\left(a+a^\dagger\right)+H_{QD}+H_L+H_R+H_T
\end{equation}
written in natural units ($e=\hbar=k_b=1$). The first term represents the vibrational energy of the molecule and the second one, electron-phonon coupling ($a$ and $a^\dagger$
are the creation and annihilator operators of the phonons in the molecule).\newline

This Hamiltonian can be simplified using the \emph{Lang-Firsov} unitary transformation \cite{LangFirsov}:
\begin{eqnarray}
	\label{LF}
	\tilde{H}&=&S\;H\;S^\dagger\nonumber\\
	S&\equiv& e^{g\;d^\dagger\;d\;\left(a-a^\dagger\right)}
\end{eqnarray}

Using this transformation, the coupling term between phonons and electrons can be simplified by choosing the parameter as $g=\frac{\lambda}{\omega_0}$. Then, the transformed Hamiltonian is
\begin{eqnarray}
  \label{H_AH}
  \tilde{H}_0&=&\left(\epsilon_0-\frac{\lambda^2}{\omega_0}\right)\tilde{d}^\dagger\;\tilde{d}+\hbar\;\omega_0\;\tilde{a}^\dagger\tilde{a}+H_L+H_R\nonumber\\
  \tilde{H}_T&=&\sum_{\textbf{k},\mu=L,R}V_{\textbf{k},\mu}\left(c_{\textbf{k},\mu}^{\dagger}\;\tilde{d}+\tilde{d}^\dagger\;c_{\textbf{k},\mu}\right)
\end{eqnarray}
defining $\tilde{\epsilon}\equiv (\epsilon_0-\lambda^2/\omega_0)$. Then, the transformed fermionic and bosonic operators can be written in terms of the original ones:
\begin{eqnarray}
	\tilde{d}&=&e^{g\left(a-a^\dagger\right)}d\nonumber\\
	\tilde{a}&=&a+g\;d^\dagger\;d
\end{eqnarray}

Due to this transformation, bosonic and fermionic degrees of freedom are mixed. As it will be shown, the effect of the coupling term between electrons and phonons is the appearance of replicas
 of the level of the original dot level at multiples of the phonon frequencies (figure \ref{diagramaAH} b).\newline

The average occupation number of the dot can be computed as:
\begin{equation}
  \label{n_d}
	n_d=\frac{1}{\pi}\int{d\omega\; \mbox{Im}\left[G^{r}(\omega;n_d)\right]\;\left[\frac{\Gamma_L\;f_L(\omega)+\Gamma_R\;f_R(\omega)}{\Gamma_L+\Gamma_R}\right]}
\end{equation}
where $f_L$ and $f_R$ are the Fermi distribution function of the left and right electrodes. Finally, it can be demonstrated that the current can be computed in this model using spectral function 
(\ref{espectral})
\begin{equation}
 \label{intensidad}
 I(V)=\int \frac{d\omega}{2\pi}\frac{\Gamma_R\;f_R(\omega)-\Gamma_L\;f_L(\omega)}{\left(\Gamma_L+\Gamma_R\right)}A(\omega)
\end{equation}
being $A(\omega)$ the local spectral function at the dot (\ref{espectral}).\newline

In the next pages, the effects of the vibrational degree of freedom are going to be studied. The problem has no exact solution (some of the Feynman 
diagrams of the exact series are represented in figure \ref{total_pert}). Therefore, it is necessary to find a good approximation to describe the behavior of the system. The first and most simple 
approximation that can be used is the so atomic limit, corresponding to the limit $\Gamma  \rightarrow 0$. In this limit the Green's function can be computed exactly and it is the same than the zero 
order in perturbation expansion (first diagram of figure \ref{total_pert}). One function interesting for studying the system is the polaron correlator, defined as:

\begin{figure}
  \centering
	\includegraphics[width=1.0\textwidth]{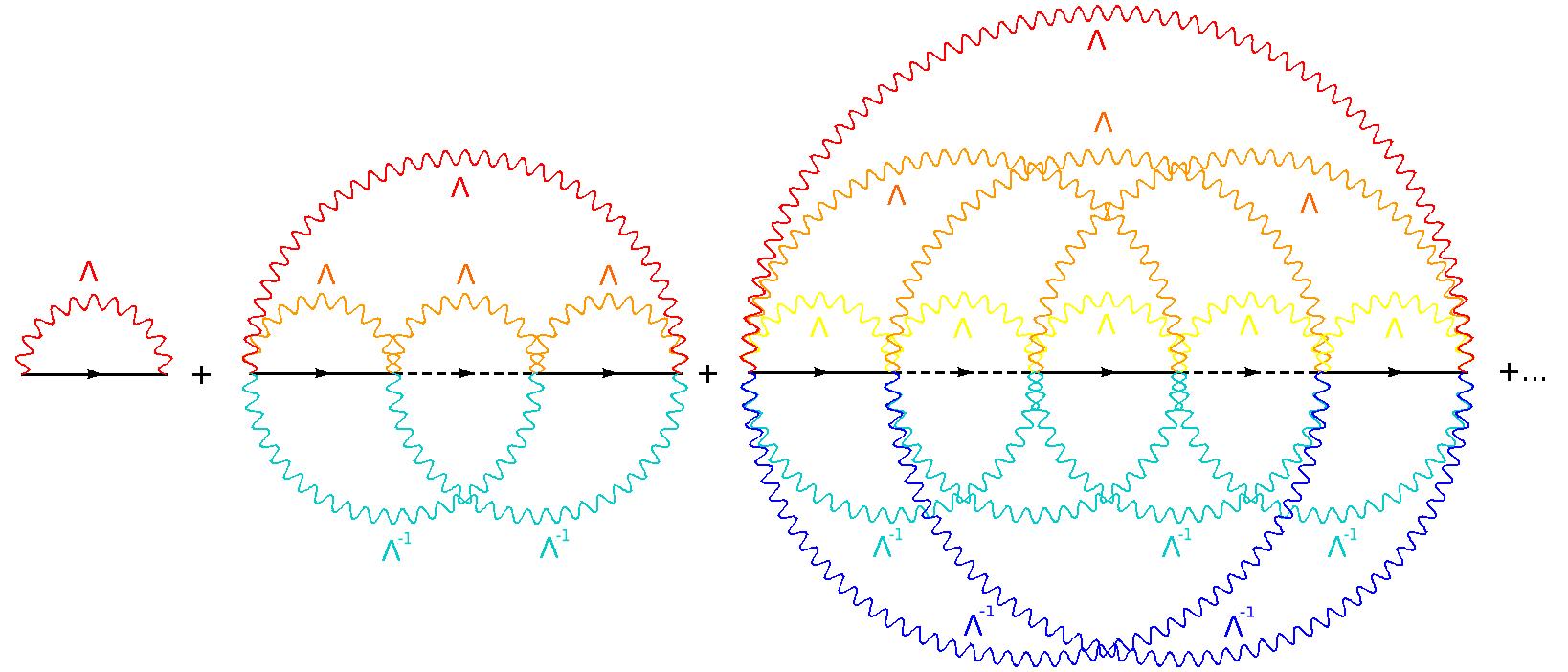}
  \caption{\small{Some diagrams from the exact perturbation series for the exact Green's function. Continuous horizontal line represents the Green's function of the electron in the molecule
 ($G_0$), the dashed one the correlation function of the electrodes ($g_k$) and wavy lines the polaron correlator (\ref{polaron}). In this diagrams correlators connecting hopping events in the same 
\emph{sense} correspond to the inverse function $\Lambda^{-1}$.}}
  \label{total_pert}
\end{figure}
\begin{eqnarray}
	\label{polaron}
	\Lambda^{+-}(t,t')=\left\langle e^{g\left(a(t)-a^\dagger(t)\right)}e^{-g\left(a(t')-a^\dagger(t')\right)}\right\rangle=\nonumber\\
\exp\left\{g^2\left[(1+n_p)\;\left(e^{-i\;\omega_0\;(t-t')}-1\right)+n_p\;\left(e^{-i\;\omega_0\;(t-t')}-1\right)\right]\right\}
\end{eqnarray}

where $n_p$ is the average population of phonons, given by Bose-Einstein distribution. Finally, the Green's function of the system in the atomic limit can be written in frequency domain, as:
\begin{equation}
  \label{gcc(w)}
 G^{R}_{0}(\omega)=\sum_{k=-\infty}^{\infty}{\left(\frac{\alpha_k\;n_d+\alpha_{-k}\;(1-n_d)}{\omega-\tilde{\epsilon}-k\;\omega_0+i\eta}\right)}
\end{equation}
where $eta$ is an infinitesimal and $\alpha$ coefficient at finite temperature is defined as
\begin{equation}
 \alpha_k=e^{-g^2(2\;n_p+1)}\;I_k\left(2g^2\sqrt{n_p(1+n_p)}\right)\;e^{k\beta\omega_0/2}
\label{alpha}
\end{equation}
Where $I_k$ is the modified Bessel function of the first kind, which is symmetric in $k$ argument ($I_k=I_{-k}$). At zero temperature it can be written as
\begin{equation}
 \alpha_k =\left\{\begin{array}{rrr}
e^{-g^2}\frac{g^{2k}}{k!}&\mbox{if }k\geq0\\
0&\mbox{if }k<0
\end{array}\right.
\end{equation}

\section{Current intensity}
In this first part of the work we will study the current intensity through molecular junction, using some existing approximations that can be used in the strong plaronic regimen.

\subsection{Polaron Tunneling Approximation (PTA):}
A first approximation that is going to be described is the so called \emph{Polaron Tunneling Approximation}, which considers that phonons are excited when electrons jump to 
the molecule, and relax when they abandon it \cite{paperPTA}. This is a good approximation when phonons can relax between two jumps of an electron or, in terms of 
coupling constants, if $\Gamma\ll\lambda^2/\omega_0$ ($\Gamma$ is the coupling constant of the molecule with the leads, and $\lambda$ the one for the coupling between 
electrons and phonons). With this considerations, the Green's functions can be calculated by summing the diagrams presented in figure \ref{PTA}. 
Using Dyson equation (\ref{Dyson1}) in this problem, the result is:
\begin{figure}
  \centering
	\includegraphics[width=1.0\textwidth]{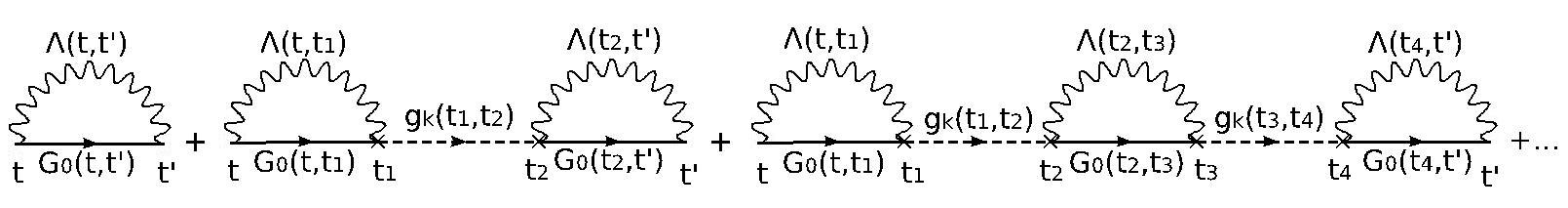}
  \caption{\small{Feynman diagrams for the PTA approximation. $G_0$ is the Green's function of the non interacting quantum dot (\ref{g0}), $\Lambda$ is the polaronic correlator and
 $g_k$ represents the Green's function of each electrode. The zero order of this series corresponds with the coupling term between the atomic limit of equation (\ref{gcc(w)})}}
  \label{PTA}
\end{figure}
\begin{eqnarray}
	G^{R}_{PTA}(\omega)=\frac{G^{R}_{0}(\omega)}{1+i\;(\Gamma_L+\Gamma_R)G^{R}_0(\omega)}
\end{eqnarray}
This means that the retarded Green's function depends on the average occupation number, which is determined using the Green's function (\ref{n_d}). The way to solve numerically this integral equation is, starting
from a given value of occupation number, compute $n_d$ using last integral and using it again as an input. This loop is repeated until convergence is reached.\newline

This simple approximation has demonstrated to give a good description for the system for frequencies close to zero \cite{Alfredo}. However, for relatively high frequencies, the behavior of this function 
is pathological because the width of the peaks tends to zero while their height remains constant.

\subsection{Single Particle Approximation (SPA):}
\begin{figure}
  \centering
	\includegraphics[width=1.0\textwidth]{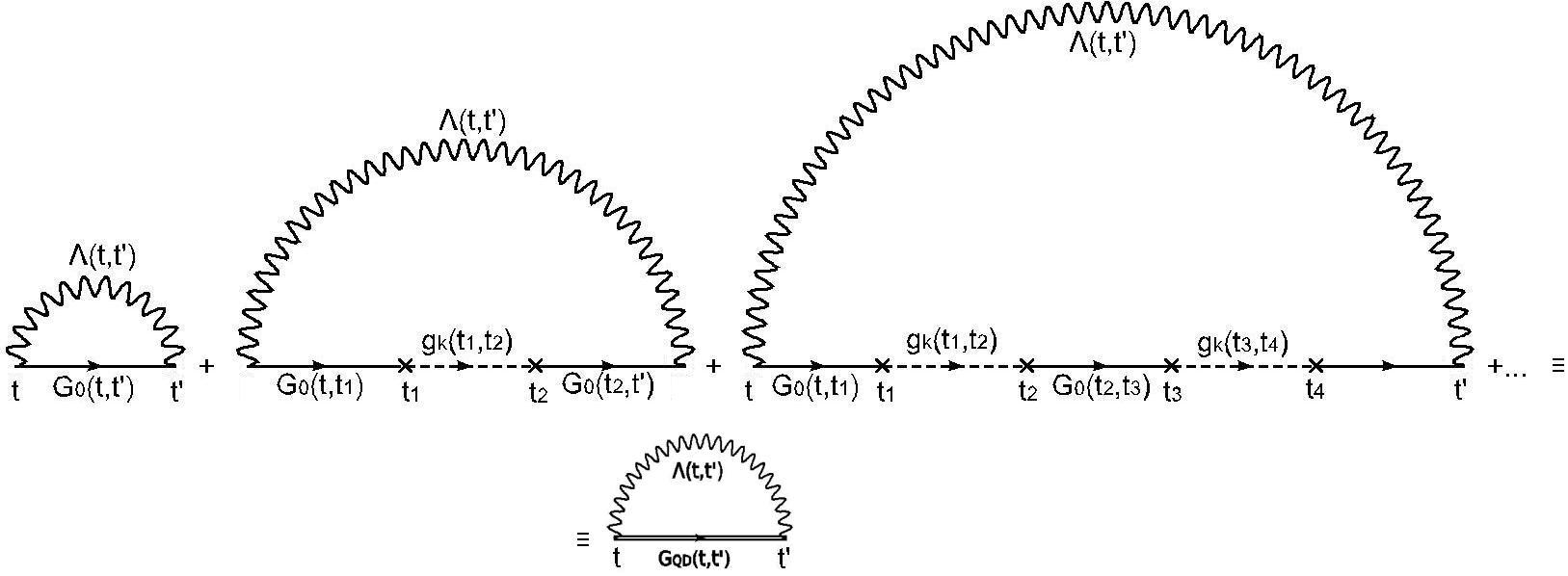}
  \caption{\small{Series of Feynman diagrams for SPA approximation. $G_0$ is the Green's function of isolated quantum dot (\ref{g0}), $\Lambda$ 
is the polaronic correlator and $g_k$ represent the Green's function of the electrodes ($k=L,R$).}}
  \label{SPA}
\end{figure}

In this approximation, the bosonic and fermionic operators are directly decoupled \cite{Flensberg}. Written in terms of Feynman diagrams, this corresponds to sum the 
perturbation series described in figure \ref{SPA}. The Greens function of the quantum dot connected to the leads is written in (\ref{g0}). Finally, for dressing the Green's function 
with the polaron correlator (\ref{polaron}), next equality can be used:
\begin{equation}
	\label{convolution}
	G^{\alpha\beta}_{SPA}(t,t')=G^{\alpha\beta}_{QD}(t,t')\;\Lambda^{\alpha\beta}(t,t')\Rightarrow G^{\alpha\beta}_{SPA}(\omega)=G^{\alpha\beta}_{QD}(\omega)\otimes\Lambda^{\alpha\beta}(\omega)
\end{equation}
where the $\otimes$ represents the convolution product of both functions. $\alpha\beta$ are the Keldysh contour indexes, which denote the branch where the time argument is. The retarded component can 
be computed exactly by using equation (\ref{retarded-keldysh}). Its imaginary part, which can be used to 
calculate the intensity (\ref{intensidad}), can be written as:
\begin{eqnarray}
	\mbox{Im}\left[G^{r}_{SPA}\right]= -\sum_{k=-\infty}^{\infty}\sum_{\mu=L,R}\Gamma_\mu\left(\frac{\alpha_k\;f_\mu(\omega+k\omega_0)+\alpha_{-k}\left(f_\mu(\omega+k\omega_0)-1\right)}{(\omega+k \omega_0-\tilde{\epsilon})^2+\Gamma^2}\right)
\end{eqnarray}

A similar approximation is obtained by using the Green's function of the atomic limit (\ref{gcc(w)}), but broadening all the poles by $\Gamma$ \cite{Flensberg}.
Using this approximation, Green's function can be written as follows:
\begin{equation}
 G^{R}_{A-SPA}=\sum_{k=-\infty}^{\infty}\left(\frac{\alpha_k\;n_d+\alpha_{-k}\;(1-n_d)}{\omega+k\;\omega_0-\tilde{\epsilon}+i\Gamma}\right)
\end{equation}

Both SPA approximations give similar results. In general, this idea describes the behavior of the system at relatively high frequencies and voltages. Moreover, this approximation is exact in the case
when the occupation number is $1$ or $0$ (full or empty level). However, 
it doesn't describes with a good agreement the slope of the current at low voltage with a good accuracy.\newline

Summarizing, the simple existing approximations for describing this system give good results at different regimes, but none of them are correct for the whole range of parameters and frequencies. 
The aim of this work is to find a simple approximation that gives good results at every frequency and voltage.

\subsection{Dressed Tunneling Approximation:}
In the literature, there are some other approximations where more complete diagrammatic resummations are considered \cite{Zazunov}. However, these approximations are usually quite complex, or 
can be used only in a certain range of parameters. The aim of this work is to find another approximation that can be used for a wide range of voltages and values of the electron-phonon coupling.\newline

\begin{figure}
  \centering
	\includegraphics[width=1.0\textwidth]{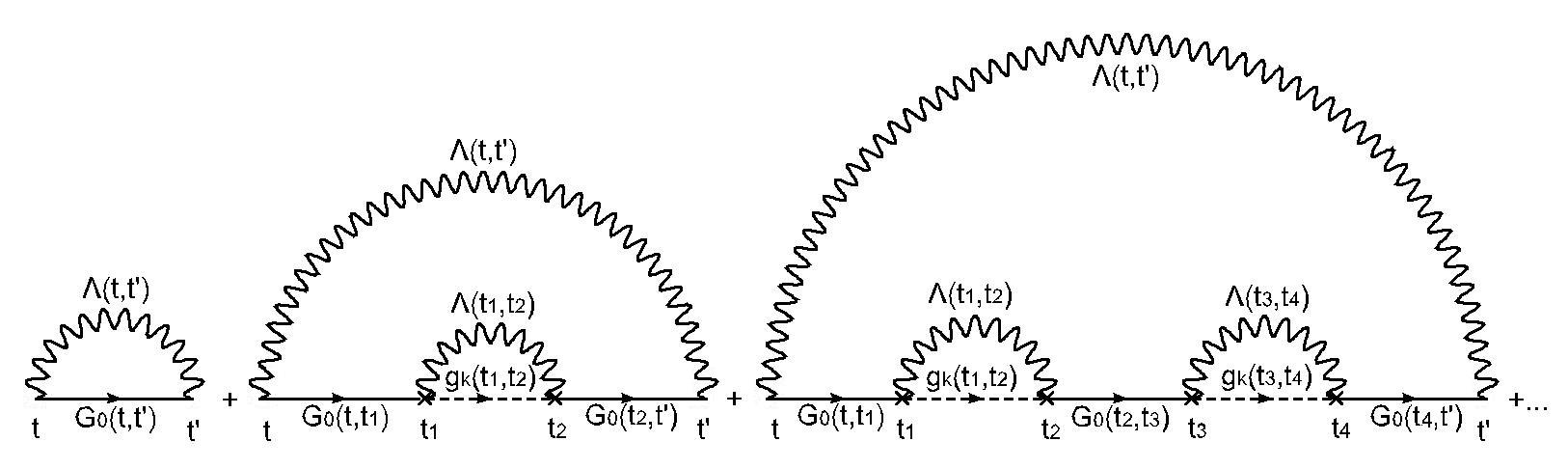}
  \caption{\small{Feynman diagrams for the DTA approximation. $G_0$ is the Green's function of the isolated quantum dot (\ref{g0}), $\Lambda$ is the polaronic correlator
 and $g_k$ represents the Green's function of the leads ($k=L,R$).}}
  \label{MSPA}
\end{figure}

The starting point for solving the problem is to consider the exact series of Feynman's diagram, represented in part in figure \ref{total_pert}. In this approximation, phonons cannot relax
from one electronic jump to another (the average lifetime of the phonons is much higher than the average lifetime of the electronic states in the molecule). This means that the Green's function in 
time domain will have  have terms with the form  $\Lambda(t_i,t_{i+1})\;\Lambda^{-1}(t_i,t_{i+2})$, being $t_{i+2}$ very close to $t_{i+1}$. 
These kind of terms are going to be simplified, in order to build our approximation. In figure \ref{MSPA}, the Feynman diagrams of this scheme are presented.\newline

Mathematically, the series can be computed by dressing the self-energy with the polaronic correlator, summing the resulting series over all possible diagrams (using Dyson equation) and dressing for 
a second time the whole Green's function. The Keldysh components of the dressed self energy can be computed using $\Sigma^{\alpha\beta}=\Sigma^{\alpha\beta}_0\otimes\Lambda^{\alpha\beta}$
\begin{eqnarray}
 \Sigma^{+-}&=&-2\;i \sum_{k=-\infty}^{\infty}\alpha_k\;\sum_{j=L,R}\Gamma_j f_j(\omega+k\;\omega_0)\nonumber\\
 \Sigma^{-+}&=&-2\;i \sum_{k=-\infty}^{\infty}\alpha_{-k}\sum_{j=L,R}\Gamma_j \left[ f_j(\omega-k\;\omega_0)-1\right]\nonumber
\end{eqnarray}

Now, the retarded component of the self energy can be determined in the wide band approximation as $\Sigma^R=\theta(t)\left[\Sigma^{+-}(t)-\Sigma^{-+}(t)\right]$
\begin{equation}
 \Sigma^R(\omega)\approx -i \sum_{k=-\infty}^{\infty}\sum_{j=L,R}\Gamma_j\left[\alpha_k\;f_j(\omega+k\;\omega_0)+\alpha_k\left(f_j(\omega-k\;\omega_0)-1\right)\right]
\end{equation}

The final result for the Green's function in the described approximation can be written in terms of this dressed self energy:
\begin{eqnarray}
 G^{R}_{DTA}=\sum_{k=-\infty}^{\infty}\left[\frac{\alpha_k\;\Sigma^{-+}(\omega+k\;\omega_0)-\alpha_{-k}\;\Sigma^{+-}(\omega+k\;\omega_0)}{\left|\omega-\tilde{\epsilon}-\Sigma^R (\omega+k\;\omega_0)\right|^2}\right]
\end{eqnarray}
This scheme, that we call \emph{Dressed Tunneling Approximation} (DTA), incorporates the effects of the polaronic cloud at any order in perturbation theory, in the strong electron-phonon coupling 
(polaronic regime).\newline

In a similar way as described in section \textbf{3.2} for the SPA approximation, the DTA result can be obtained approximately by broadening the poles of the atomic Green's function with the dressed 
self energy. In this case we obtain
\begin{equation}
 G^{R}_{A-DTA}=\sum_{k=-\infty}^{\infty}\left[\frac{\alpha_k\;n_d+\alpha_{-k}\;(1-n_d)}{\omega-\tilde{\epsilon}+k\;\omega_0-\Sigma^{R}(\omega)}\right]
\end{equation}

This approximation will be labeled as \emph{Atomic-limit Dressed Tunneling Approximation} (A-DTA). In the next section, the results of both approximations will be presented and compared with 
others found in the literature and with some numerical results obtained using quantum Monte Carlo calculations \cite{NMC}.

\subsection{Results}
\begin{figure}
  \centering
    \begin{minipage}{0.7\linewidth}
      \includegraphics[width=1\textwidth]{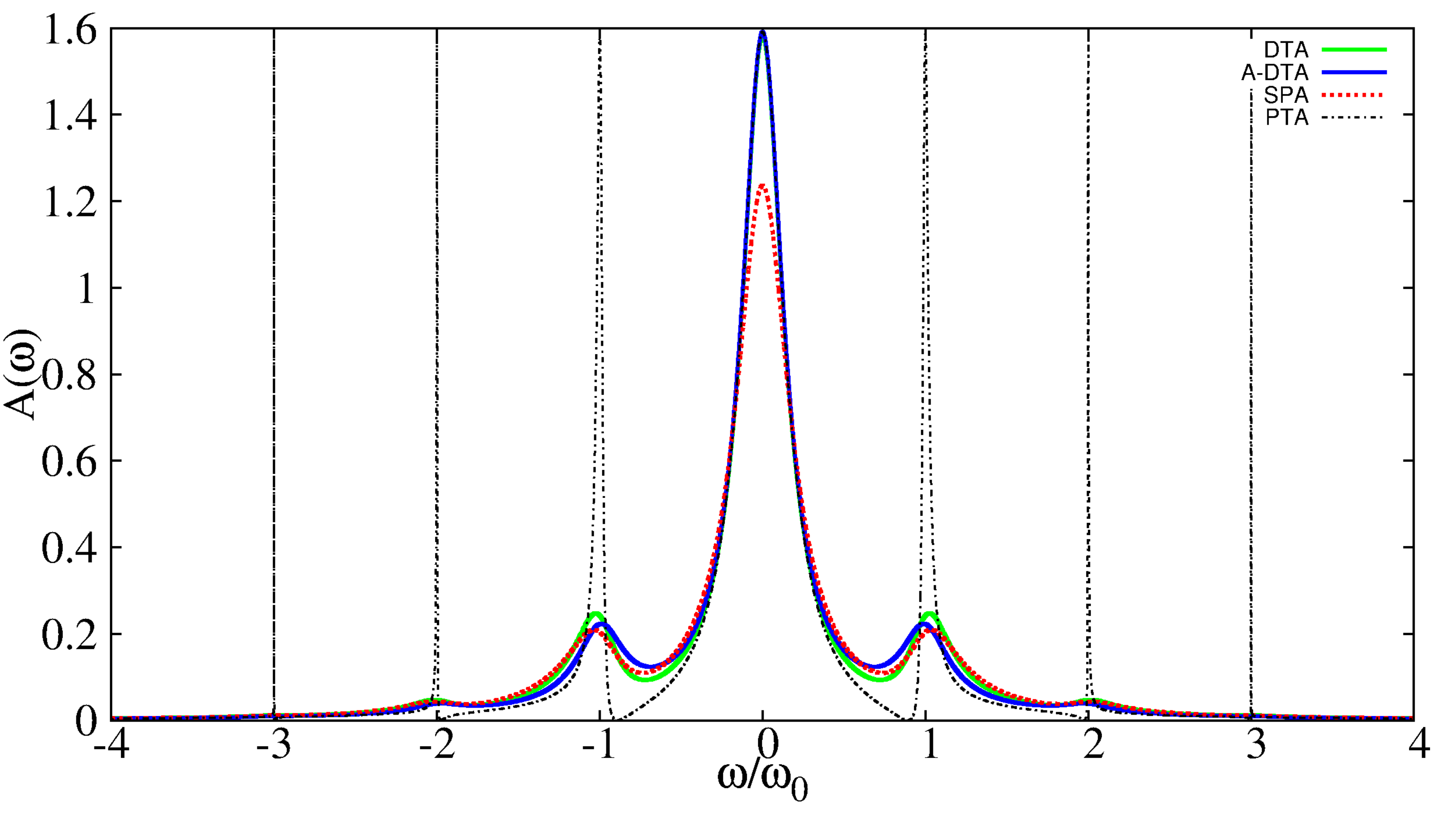}
    \end{minipage}\\
    \begin{minipage}{0.7\linewidth}
      \includegraphics[width=1\textwidth]{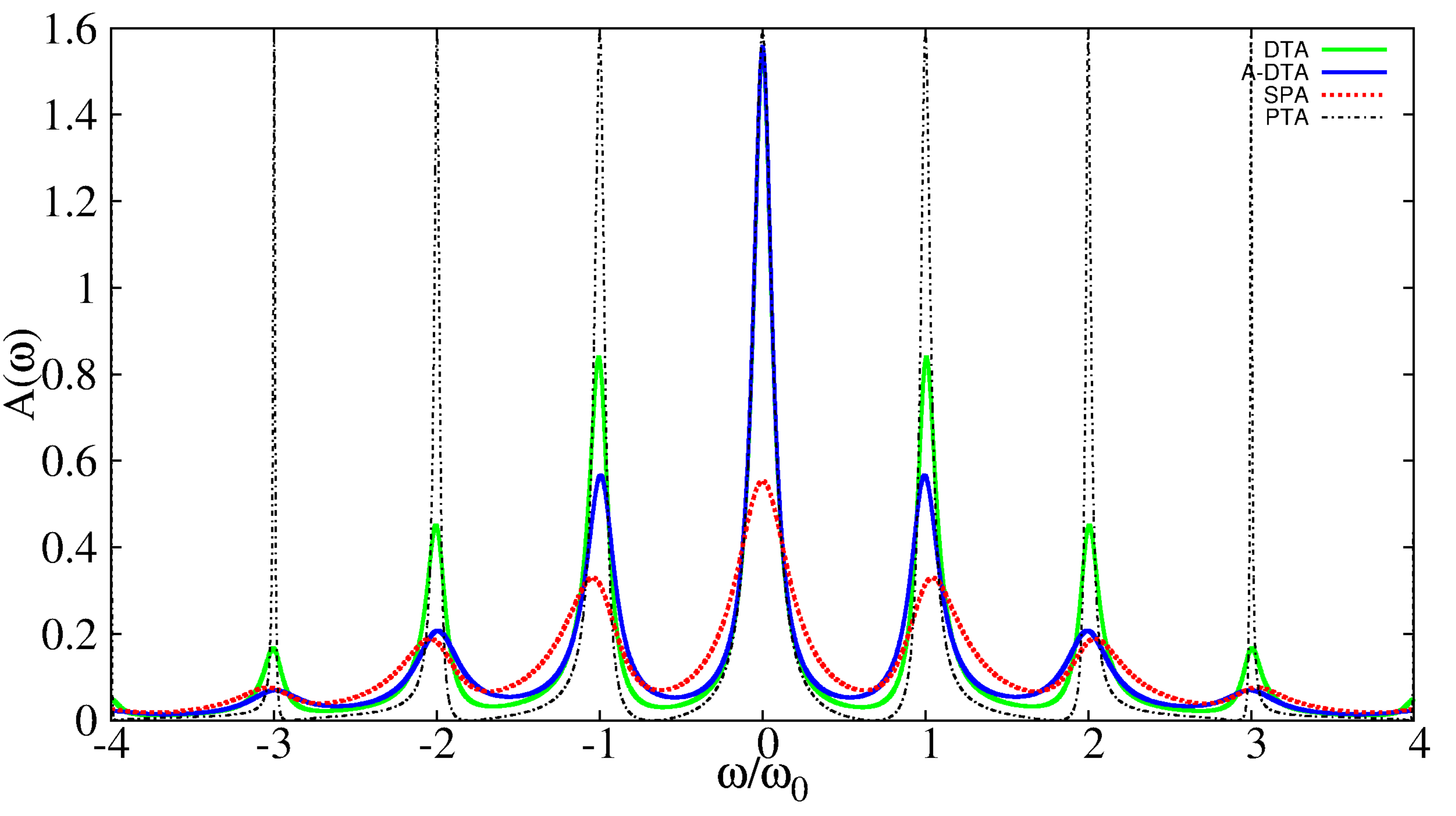}
    \end{minipage}\\
    \begin{minipage}{0.7\linewidth}
      \includegraphics[width=1\textwidth]{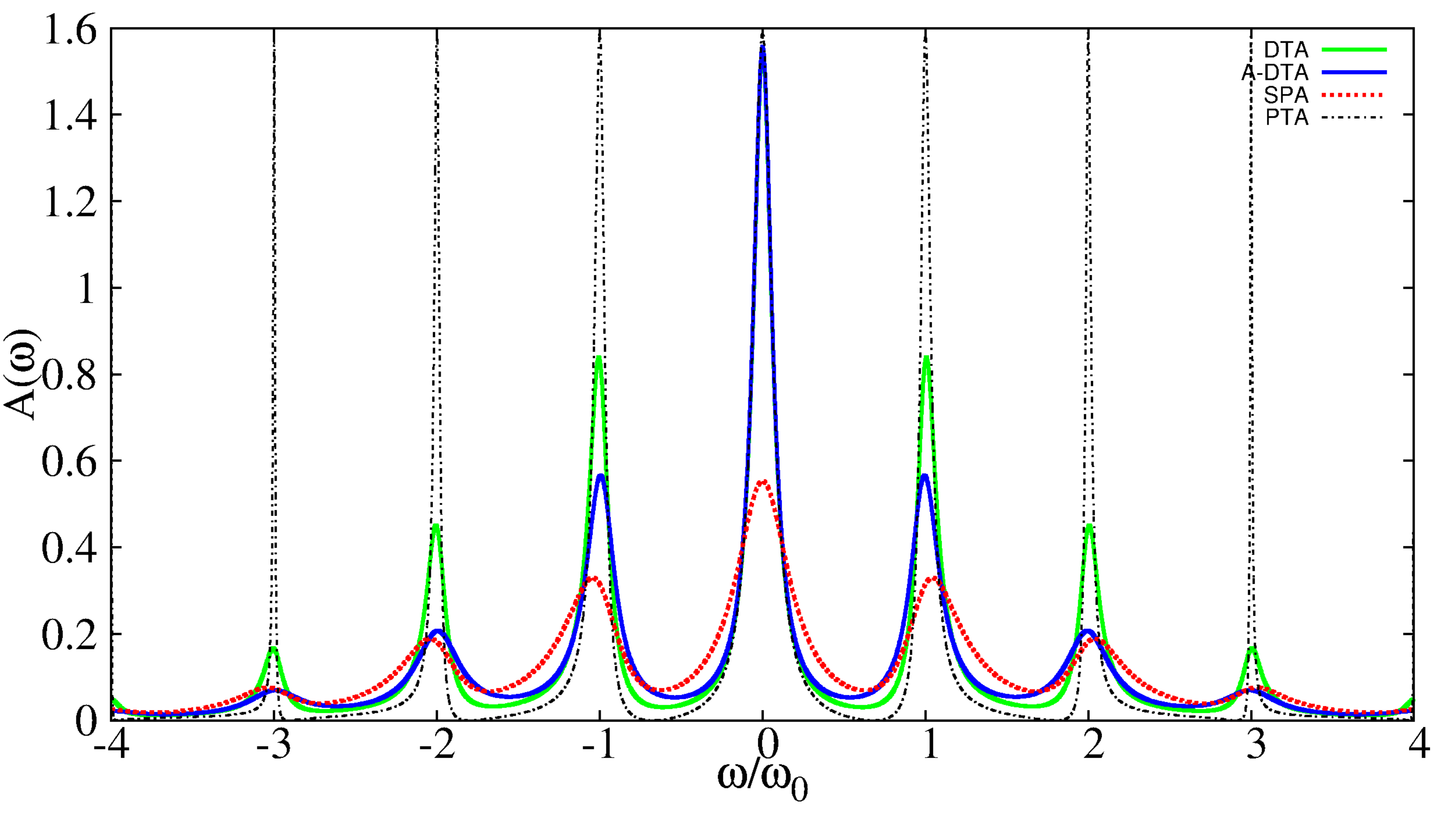}
    \end{minipage}
  \caption{\small{Spectral function for different values of coupling in the symmetric case and zero bias voltage for DTA (continuous green), A-DTA (continuous blue), SPA (discontinuous red) and 
PTA (discontinuous black). $\Gamma_L=\Gamma_R=0.1\;\omega_0$, $T=0.2\;\omega_0$ and $g=0.5,1,1.5$.}}
\label{espectralg}
\end{figure}
In this section we present the results of the scheme described before are going to be analyzed. As a reference, at voltage equal to zero, the Fermi levels of both electrodes are supposed to be aligned with the energy
level of the molecule. If a voltage is applied to the junction (non equilibrium case, $V_b\neq 0$) the Fermi level of the left electrode increases to an energy equal to $V_b/2$, and Fermi energy 
of the right electrode decreases this quantity (figure \ref{diagramaqd}).\newline

The first comparison that can be done between methods is through the density of states (\ref{espectral}). In figure \ref{espectralg} a comparison between methods is shown for different electron-phonon 
coupling constants and for a bias voltage equal to zero. The graphics presented are all symmetric with respect to the origin of frequencies. This is because the energy level of the molecule has
been considered to be aligned with equilibrium Fermi level of the electrodes (zero gate voltage). In this situation there is an extra symmetry in the system between holes and electrons so, the
energy level of the molecule has the same possibility to be empty or occupied ($n_d=1/2$). The density of states presented is composed by Lorentzian peaks and
properly normalized to, $\int d\omega\;A(\omega)=1$.\newline

As it is shown in figure \ref{espectralg}, at low frequencies, both approximations introduced in last section (DTA and A-DTA) tend to the PTA approximation. In this range we expect them to be a good
approximation for describing the behavior of the system \cite{Alfredo}. At high frequencies, both approximations tend to be similar to the SPA approximation. When electron-phonon coupling constant increases 
(a more polaronic regime), the peaks become sharper and the secondary ones start to be more important. The PTA approximation has almost no dependence on this coupling parameter and it presents a
pathological behavior (peaks always have the same height but its width tends to zero when going to higher frequencies).\newline
\begin{figure}
  \centering
    \begin{minipage}{0.7\linewidth}
      \includegraphics[width=1\textwidth]{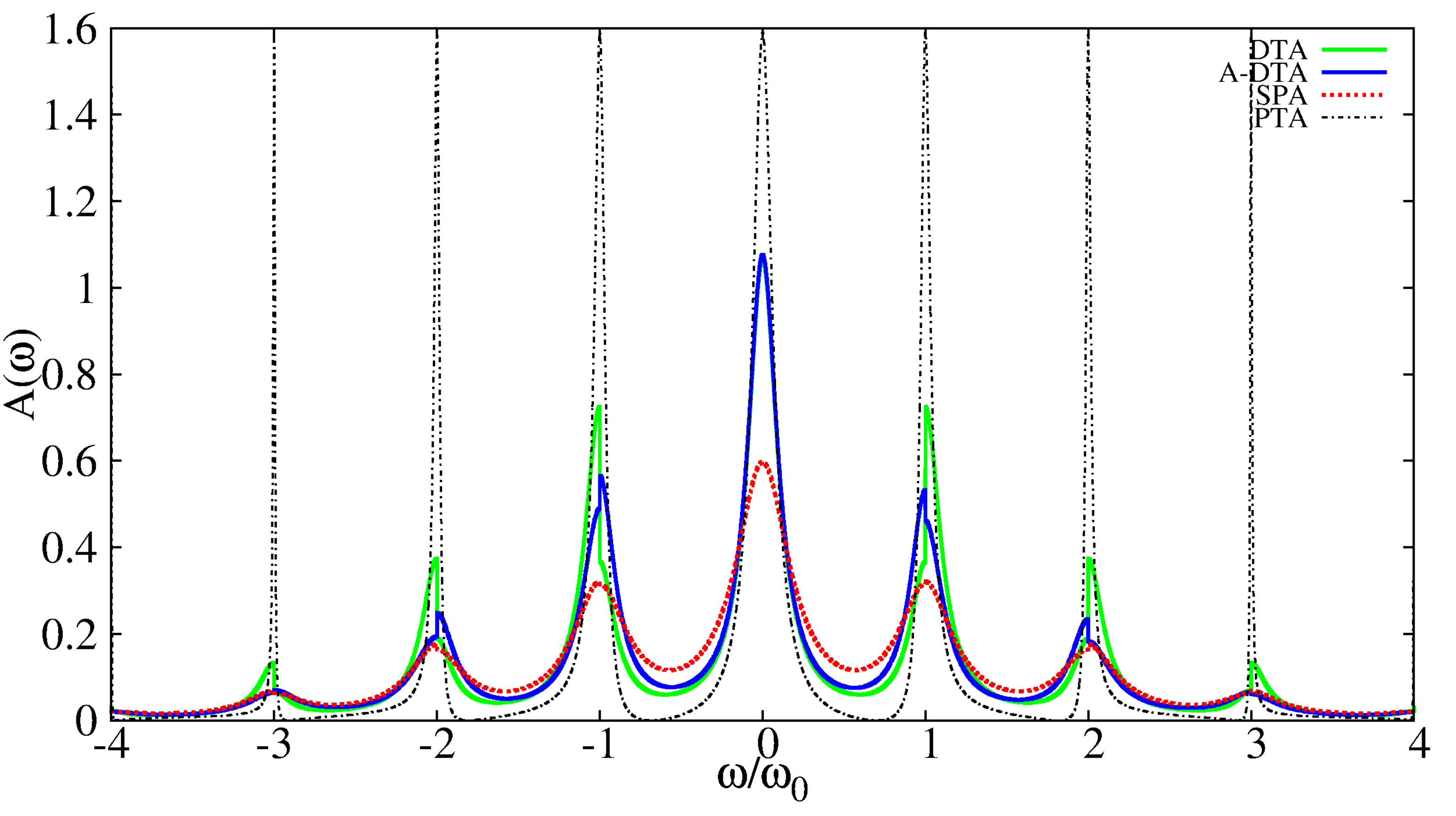}
    \end{minipage}\\
    \begin{minipage}{0.7\linewidth}
      \includegraphics[width=1\textwidth]{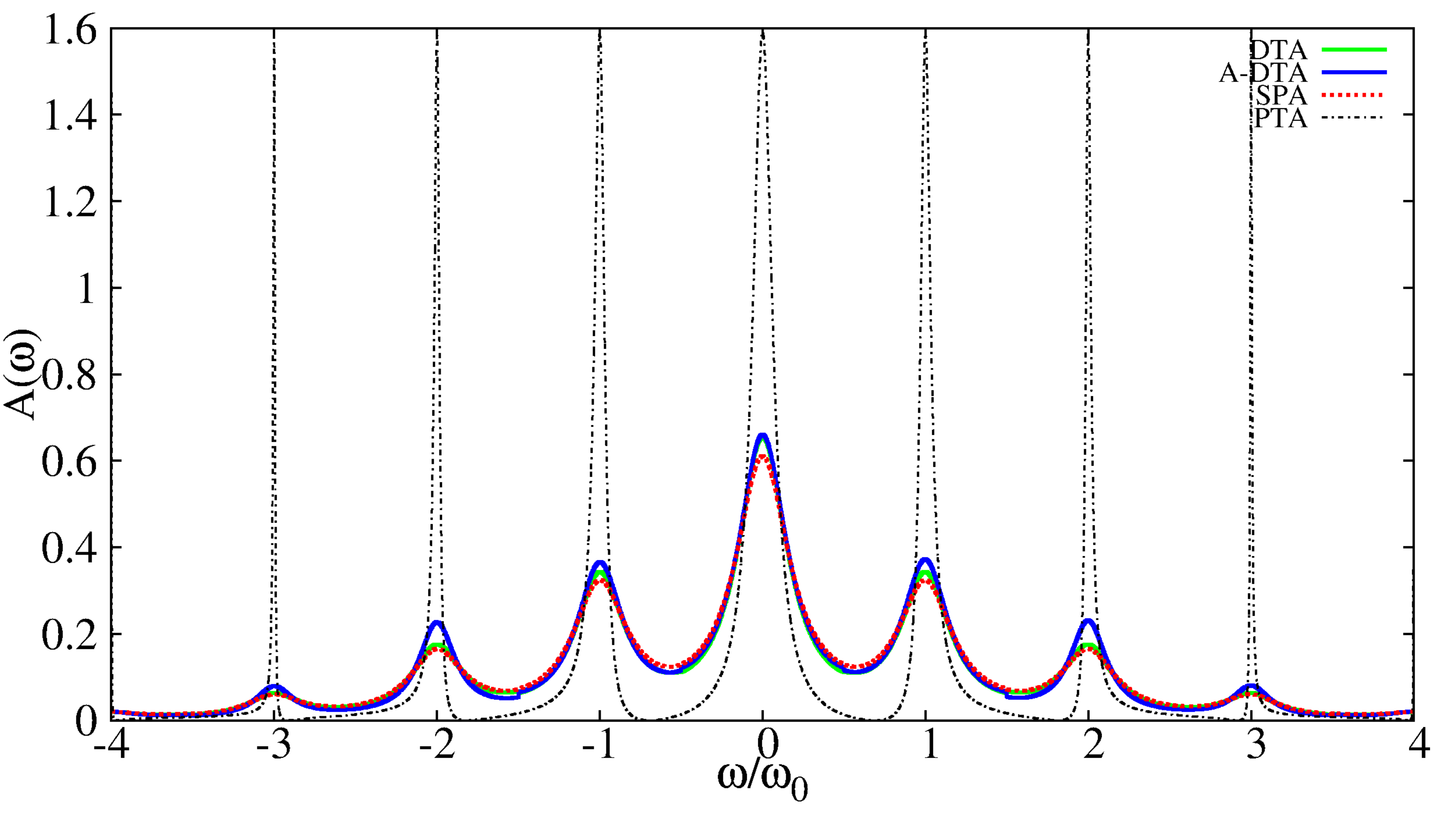}
    \end{minipage}
  \caption{\small{Spectral function for non equilibrium case for DTA (continuous green) A-DTA (continuous blue), SPA (discontinuous red) and 
PTA (discontinuous black) approximations. Parameters of the model used for this plots are: $\Gamma_L=\Gamma_R=0.1\;\omega0$, $T=0.2\;\omega_0$, $g=1$ and $V_b=2,5\;\omega_0$.}}
\label{espectralv}
\end{figure}

One important observation is that, even in a non polaronic regime, both \emph{DTA} approximations provide good results in the two known limits (zero and high frequencies). So this approximation is 
expected to be valid at any range of electron-phonon coupling. This figure was for a bias voltage equal to zero, meaning that it is an equilibrium case. Qualitatively, this approximation can also be 
compared with calculations using \emph{scattering states numerical renormalization group} \cite{SNRG}. We have verified that the central peak has the same height, and the position of the 
other ones are reproduced with a good accuracy.\newline 

 In figure \ref{espectralv} a non equilibrium case is shown for two different values of bias voltage. This figures have been computed considering a stationary case. This means that the system is 
supposed to relax after applying the bias voltage to the junction. As happens at high frequencies, when the voltage increases the approximations described tend to be similar to the SPA one. 
This means that at high energies will tend to behave as if the electrons and phonons were uncoupled. In these conditions electrons will not feel intensely the influence of the internal degree of 
freedom of the molecule. This same behavior has been found, but not shown, when the level is displaced from the symmetric case ($\tilde{\epsilon}\neq 0$).\newline

\begin{figure}
  \centering
    \begin{minipage}{0.7\linewidth}
      \includegraphics[width=1\textwidth]{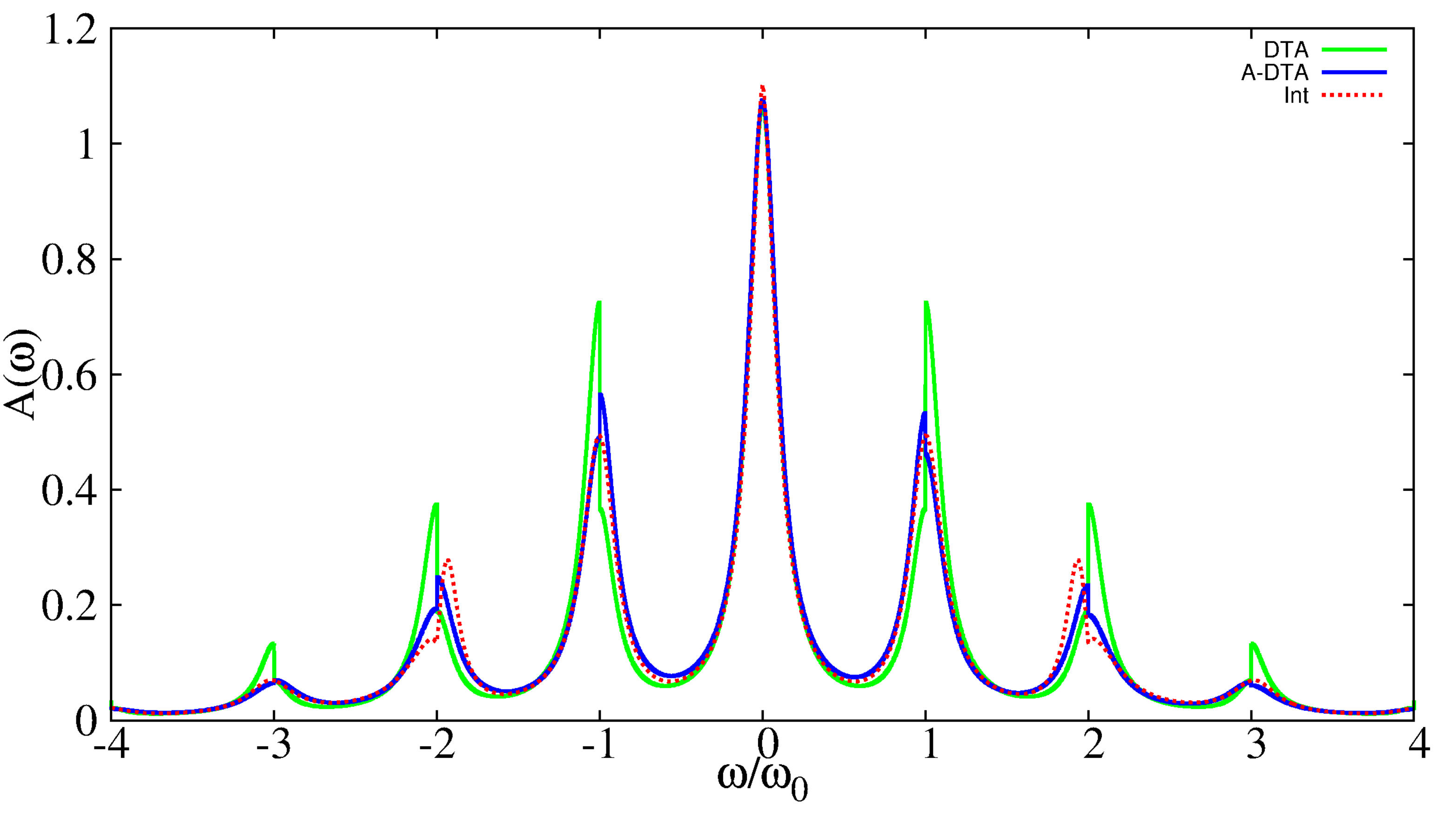}
    \end{minipage}\\
    \begin{minipage}{0.7\linewidth}
      \includegraphics[width=1\textwidth]{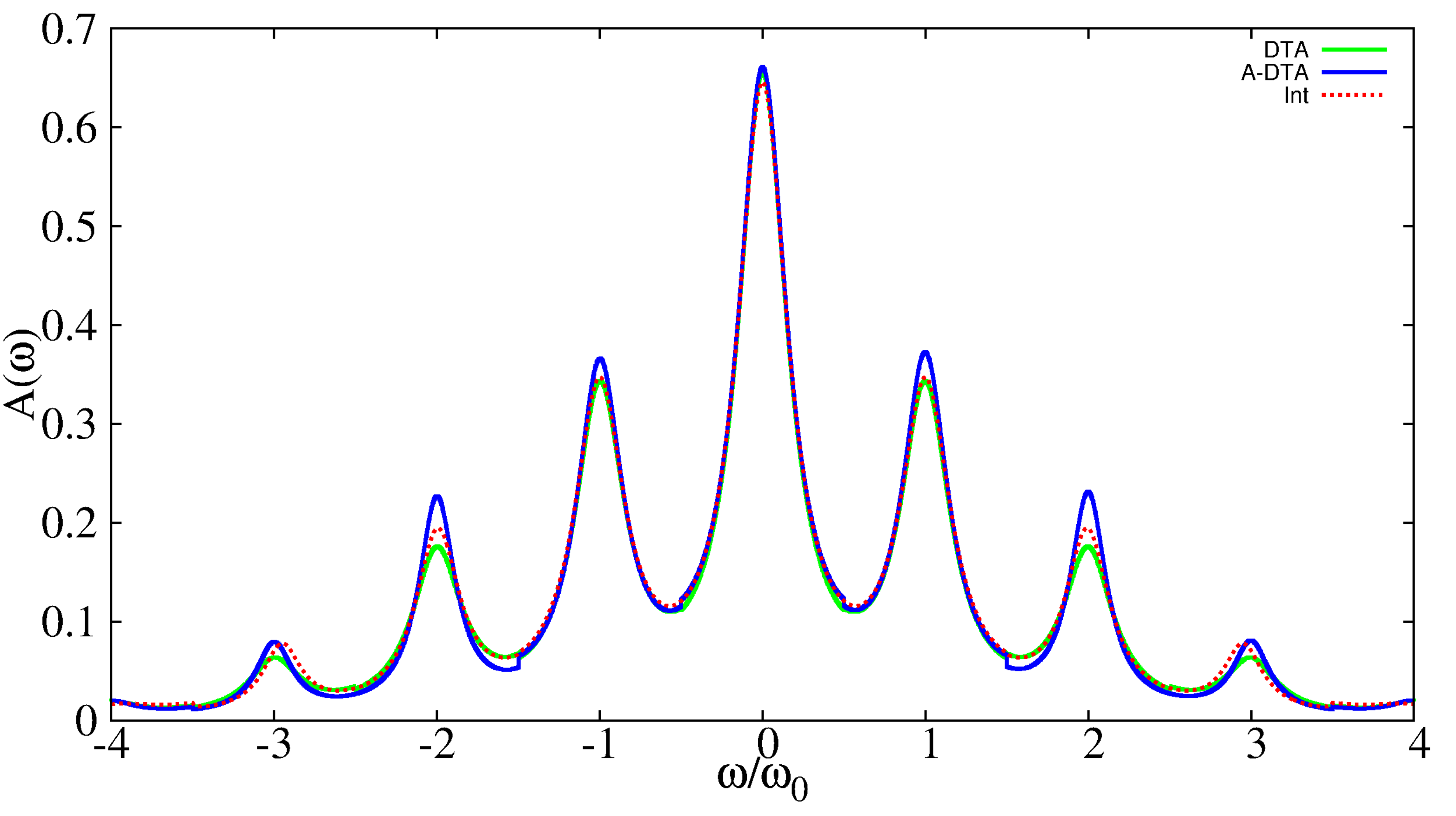}
    \end{minipage}
\caption{\small{Spectral function for non equilibrium case for DTA (continuous green) A-DTA (continuous blue) and interpolation (discontinuous orange) approximations. The parameters used in this
 plots are: $\Gamma_L=\Gamma_R=0.1\;\omega_0$, $T=0$, $g=1$ and $V_b=2,5\;\omega_0$.}}
\label{espectralvint}
\end{figure}

However, this approximation is not expected to be valid at very high values of the bias voltage ($V_b$). This is because, in this conditions electrons are expected to tunnel to the molecule with a 
higher rate, exciting real phonons. If the current is high enough, the phonons distribution starts to deviate form an equilibrium one (given by the Bose-Einstein distribution at a given temperature). To
solve this problem it would be necessary to consider non equilibrium Green's functions also for the phonons, and to find an appropriate approximation for this functions.\newline

Our approximation can be also be compared with the one described in \cite{int} where the authors use an interpolative self-energy approximation. In figure \ref{espectralvint} we show both 
\emph{DTA} approximations and the interpolative self-energy one for a non equilibrium case. As it is shown in this figure, the agreement between the approximations is quite good for every frequency.\newline

\begin{figure}
  \centering
	\includegraphics[width=1\textwidth]{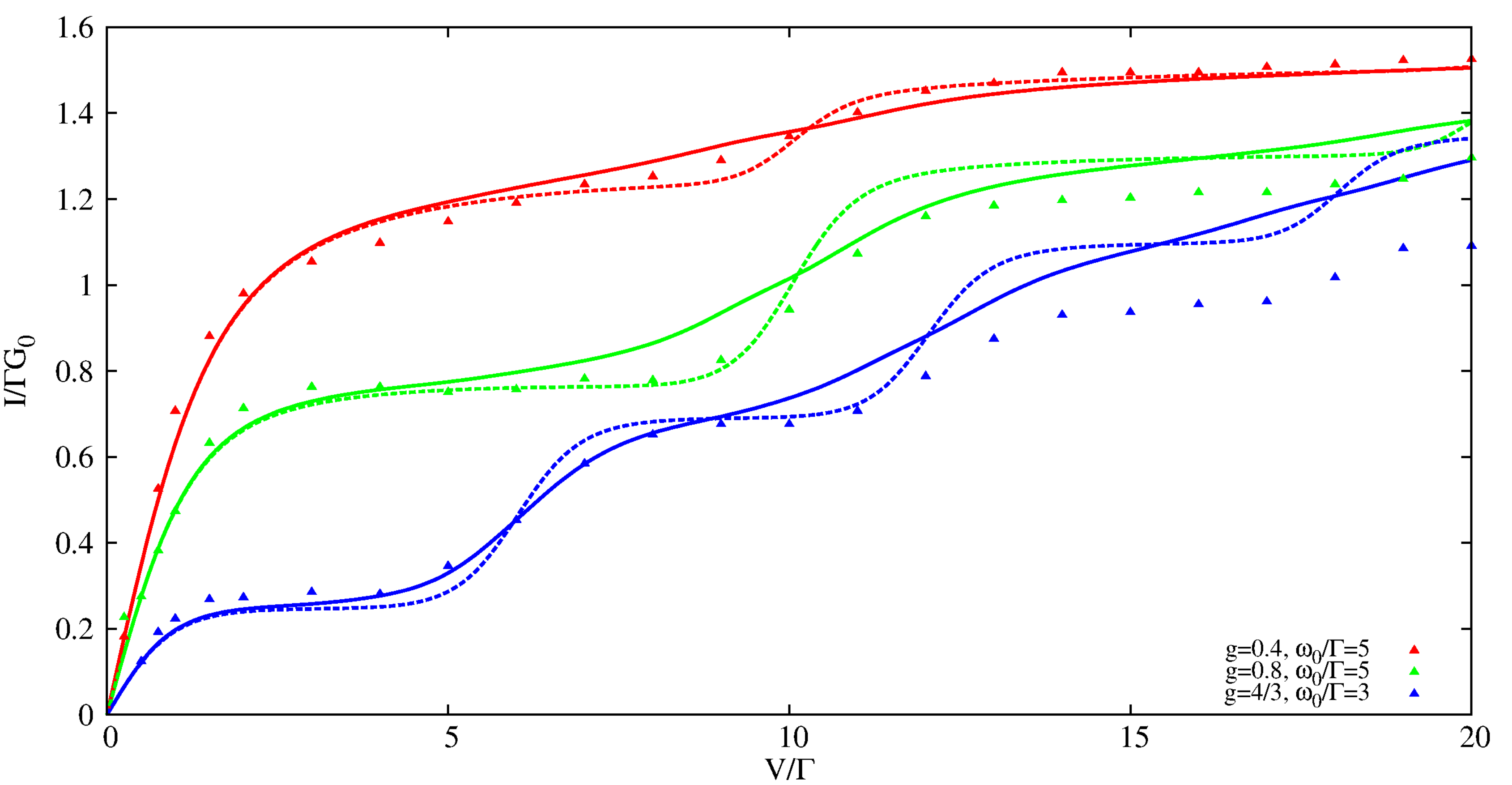}
  \caption{\small{Current for DTA (continuous line) and PTA (discontinuous line) approximation, together with path integral Monte Carlo numerical calculations (triangles) 
for a symmetric case ($\tilde{\epsilon}=0$) at $T=0.2/\Gamma$.}}
  \label{intMC}
\end{figure}

As it has been explained before, the current can be computed by integrating the spectral function multiplied by the Fermi functions, as described in the expression (\ref{intensidad}). In figure 
\ref{intMC} the current of the scheme described before is compared with the PTA approximation. We also show some numerical data presented in reference \cite{NMC}, calculated using path integral
Monte Carlo method. In this reference the authors compute the current in a symmetric and non equilibrium case, for different values of the parameters.\newline

As it is shown in the figure \ref{intMC}, the approximation proposed in this thesis provide good results for strong, but also for weak electron-phonon coupling. The slope of the current (conductance) 
of both approximations at low voltage are nearly the same. This slope is related with the height of the central peak of the density of states. Moreover, the asymptotic value
of both approximations is the same, for high voltage values.\newline

However, at high frequencies the curves start to present some deviation from the simulations in strong coupling regime. This can be due to the consideration of an equilibrium population of phonons,
condition that can not be valid in the case when the intensity is high enough. Moreover, the steps described by this approximation are not so sharp than the ones presented by the numerical 
calculations.

\newpage

\section{Shot noise:}
Usually, it is more representative for studying a given nanoscale system the analysis of current fluctuations than the main current itself. The formalism used to compute these fluctuations is the Full Counting Statistics (FCS). In this 
formalism a phase is added to the the electrons when tunneling from the electrodes to the molecule. The phase can be added in the hopping part of the Hamiltonian as:
\begin{equation}
 H_T=\left[t_{L}\;e^{i\lambda/2}d^\dagger\;\psi_{LC}+t_{R}\;d^\dagger\;\psi_{RC}+h.c.\right]
\end{equation}
The phase has been added, for simplifying, only to the left electrode term. For practical reasons, it is more convenient to define the so-called generating function, given by 
$\chi(\lambda)=\sum_q e^{iq\lambda}P_q$, where $P_q$ is the probability of the charge q to be transferred through the system \cite{shotnoise}. In the Keldysh formalism, this function can be computed as
\begin{equation}
 \chi(\lambda)=\left\langle T_C \;e^{-i\int_c dt H_T (t)}\right\rangle
\end{equation}

All the cumulants can then be compute in terms of this function
\begin{equation}
 \left\langle\delta^n \;q\right\rangle=\left.(-i)^n\;\frac{\partial^n}{\partial\;\lambda^n}\mbox{U}(\lambda)\right|_{\lambda=0}
\end{equation}
where $\mbox{U}(\lambda)=\ln\left[\chi(\lambda)\right]$. The first cumulant in this theory ($n=1$) corresponds with the average current that flows through the molecular junction. The current
flowing from the left electrode to the molecule is the same than the one going from the molecule to the other electrode. So, the formula for the average current can be simplified as
\begin{equation}
 I=\int\frac{d\omega}{2\pi}\frac{\Gamma_L\;I_R-\Gamma_R\;I_L}{\Gamma_L+\Gamma_R}=\int\frac{d\omega}{2\pi}\frac{\Gamma_L\;f_L(\omega)-\Gamma_R\;f_R(\omega)}{\Gamma_L+\Gamma_R}\mbox{Im}\left[G^{a}_{CC}(\omega)\right]
  \label{generalI}
\end{equation}

In this work also the second cumulant (shot noise) is going to be analyzed and compared with the other approximations
\begin{equation}
 S=\left.\frac{\partial^2}{\partial\lambda^2}U(\lambda)\right|_{\lambda=0}=t_{RC}^2\int{\frac{d\omega}{2\pi} \left[\left.\frac{\partial\;G_{CC}^{-+}}{\partial\lambda}\right|_{\lambda=0} g_{R}^{+-}-\left.\frac{\partial\;G_{CC}^{+-}}{\partial\lambda}\right|_{\lambda=0} g_{R}^{-+}\right]}
  \label{generalnoise}
\end{equation}

In the next sections the shot noise for every approximation described before are going to be computed.

\subsection{Polaron Tunneling Approximation}
In this approximation, described by the Feynman diagrammatic series presented in figure \ref{PTA}, The Green's function of the isolated molecule is dressed with the polaron. By summing the series and
integrating with respect to the phase $\lambda$, we find next result
\begin{equation}
 U(\lambda)=\int \frac{d\omega}{2\pi}\ln\{1+T(\omega)\left[\left(e^{i\lambda}-1\right)\;f_L(\omega)\left(1-f_R(\omega)\right)+\left(e^{-i\lambda}-1\right)\;f_R(\omega)\left(1-f_L(\omega)\right)\right]\}
\end{equation}
Where the transmission can be computed as $T(\omega)=4\Gamma_L\;\Gamma_R/\left(f^{-2}(\omega)+\Gamma^2\right)$, being
\begin{equation}
 f\equiv \sum^{\infty}_{k=-\infty}\alpha_k\;\left(\frac{n_0}{\omega-\tilde{\epsilon}+k\;\omega_0}+\frac{(1-n_0)}{\omega-\tilde{\epsilon}-k\;\omega_0}\right)
\end{equation}
Where $\alpha_k$ is defined in \ref{alpha}. This formula is similar to the Levitov-Lesovik formula \cite{Levitov} for transport through a tunnel junction, with a renormalized transmission.

\subsection{Single Particle Approximation}
In this approximation, the Green's function of the quantum dot coupled to electrodes is dressed with the polaron. It is equivalent to decouple the bosonic and fermionic degrees of freedom. The 
diagrammatic series of this approximation is presented in figure \ref{SPA}. By summing all the diagrams, the Keldysh components of the molecule needed to describe the transport phenomena 
(\ref{generalI})(\ref{generalnoise}) are
\begin{eqnarray}
  \tilde{G}_{SPA}^{+-}=-2i\sum^{\infty}_{k=-\infty}\alpha_k \frac{e^{i\lambda}\Gamma_L\;f_L(\omega+k\;\omega_0)+\Gamma_R\;f_R(\omega+k\;\omega_0)}{det(\omega+k\;\omega_0)}\nonumber\\
  \tilde{G}_{SPA}^{-+}=-2i\sum^{\infty}_{k=-\infty}\alpha_k \frac{e^{-i\lambda}\Gamma_L\;\left[f_L(\omega-k\;\omega_0)-1\right]+\Gamma_R\;\left[f_R(\omega-k\;\omega_0)-1\right]}{det(\omega-k\;\omega_0)}
\end{eqnarray}

where
\begin{equation}
 det(\omega)=(\omega-\tilde{\epsilon})^2+\Gamma^2-4\Gamma_L\Gamma_R\left[\left(e^{i\lambda}-1\right)\;f_L(\omega)(f_R(\omega)-1)+\left(e^{-i\lambda}-1\right)\;f_R(\omega)(f_L(\omega)-1)\right]
\end{equation}

With these functions, and using equation (\ref{generalnoise}), the general formula for the shot noise can be computed
\begin{eqnarray}
 S=4\;\Gamma_R\Gamma_L\int \frac{d\omega}{2\pi}\sum^{\infty}_{k=-\infty}\left[\frac{\left[\alpha_k (f_R(\omega)-1) f_L(\omega')+\alpha_{-k} f_R(\omega) (f_L(\omega')-1)\right]}{(\omega'-\tilde{\epsilon})^2+\Gamma^2}+\right.\nonumber\\
   \left.\frac{\partial}{\partial\;\lambda}det(\omega)\right|_{\lambda=0}\left.\sum_{\mu=L,R}\frac{\Gamma_\mu\left[\alpha_k f_\mu(\omega')(f_R(\omega)-1)-\alpha_{-k}(f_\mu(\omega')-1)f_R(\omega)\right]}{\left[(\omega'-\tilde{\epsilon})^2+\Gamma^2\right]^2}\right]
\end{eqnarray}
where
\begin{equation}
 \left.\frac{\partial}{\partial\;\lambda}det(\omega)\right|_{\lambda=0}=4\Gamma_L\Gamma_R\left[f_L(\omega')(f_R(\omega')-1)-\;f_R(\omega')(f_L(\omega')-1)\right]
\end{equation}
and $\omega'\equiv\omega+k\;\omega_0$

\subsection{Dressed Tunneling Approximation}
In this approximation, the time between two consecutive tunneling processes (from the molecule to one of the electrodes, and from the electrodes to the molecules) is taken to be infinitesimally short.
This approximation is described by the diagrams in figure \ref{MSPA}. By summing over all the possible diagrams, the result found for the Keldysh components Green's functions is
\begin{eqnarray}
 G^{+-}_{DTA}=-2i\;\sum^{\infty}_{k=-\infty}\alpha_k\frac{\Gamma_L\;\tilde{g}_{L}^{+-}(\omega')\;e^{i\lambda}+\Gamma_R\;\tilde{g}_{R}^{+-}(\omega')}{det(\omega')}\nonumber\\
 G^{-+}_{DTA}=-2i\;\sum^{\infty}_{k=-\infty}\alpha_{-k}\frac{\Gamma_L\;\tilde{g}_{L}^{-+}(\omega')\;e^{-i\lambda}+\Gamma_R\;\tilde{g}_{R}^{-+}(\omega')}{det(\omega')}
\end{eqnarray}
Where $\omega'=\omega+k\;\omega_0$ and, in this case
\begin{eqnarray}
 det(\omega)=(\omega-\tilde{\epsilon})^2+2\Gamma_L\Gamma_R\left(\tilde{g}_{L}^{+-}(\omega)+\tilde{g}_{L}^{-+}(\omega)\right)\left(\tilde{g}_{R}^{+-}(\omega)+\tilde{g}_{R}^{-+}(\omega)\right)+\nonumber\\
 \Gamma_{L}^2\left(\tilde{g}_{L}^{+-}(\omega)-\tilde{g}_{L}^{-+}(\omega)\right)^2+\Gamma_{R}^2\left(\tilde{g}_{R}^{+-}(\omega)-\tilde{g}_{R}^{-+}(\omega)\right)^2-\nonumber\\
 4\Gamma_L\Gamma_R\left(\tilde{g}_{L}^{-+}(\omega)\tilde{g}_{R}^{+-}(\omega)e^{-i\lambda}+\tilde{g}_{L}^{+-}(\omega)\tilde{g}_{R}^{-+}(\omega)e^{i\lambda}\right)
\end{eqnarray}
For simplifying, we have defined next quantities:
\begin{eqnarray}
 \tilde{g}_{\mu}^{+-}&=&\sum^{\infty}_{k=-\infty}\alpha_k\;f_\mu(\omega')\nonumber\\
 \tilde{g}_{\mu}^{-+}&=&\sum^{\infty}_{k=-\infty}\alpha_{-k}\;\left(f_\mu(\omega')-1\right)
\end{eqnarray}

Again, using equation (\ref{generalnoise}), the shot noise formula can be computed as:
\begin{eqnarray}
  S=4\Gamma_L\Gamma_R\int \frac{d\omega}{2\pi}\sum^{\infty}_{k=-\infty}\left[\frac{\alpha_k\left(f_R(\omega)-1\right)\tilde{g}_{L}^{+-}(\omega')-\alpha_{-k}f_R(\omega)\tilde{g}_{L}^{-+}(\omega')}{\left. det(\omega')\right|_{\lambda=0}}+\right.\nonumber\\
  \left.2i\;\Gamma_L\Gamma_R \left[G_{DTA}^{+-}(\omega)+G_{DTA}^{-+}(\omega)\right]\frac{\left(\tilde{g}_{R}^{-+}(\omega')\tilde{g}_{L}^{+-}(\omega')-\tilde{g}_{L}^{-+}(\omega')\tilde{g}_{R}^{+-}(\omega')\right)}{\left. det(\omega')\right|_{\lambda=0}}\right]
\end{eqnarray}
where:
\begin{equation}
 \left. det(\omega')\right|_{\lambda=0}=(\omega'-\tilde{\epsilon})^2+\left[\Gamma_L\left(\tilde{g}_{L}^{+-}(\omega')-\tilde{g}_{L}^{-+}(\omega')\right)+\Gamma_R\left(\tilde{g}_{R}^{+-}(\omega')-\tilde{g}_{R}^{-+}(\omega')\right)\right]^2
\end{equation}

\subsection{Results}
\begin{figure}
  \centering
	\includegraphics[width=1\textwidth]{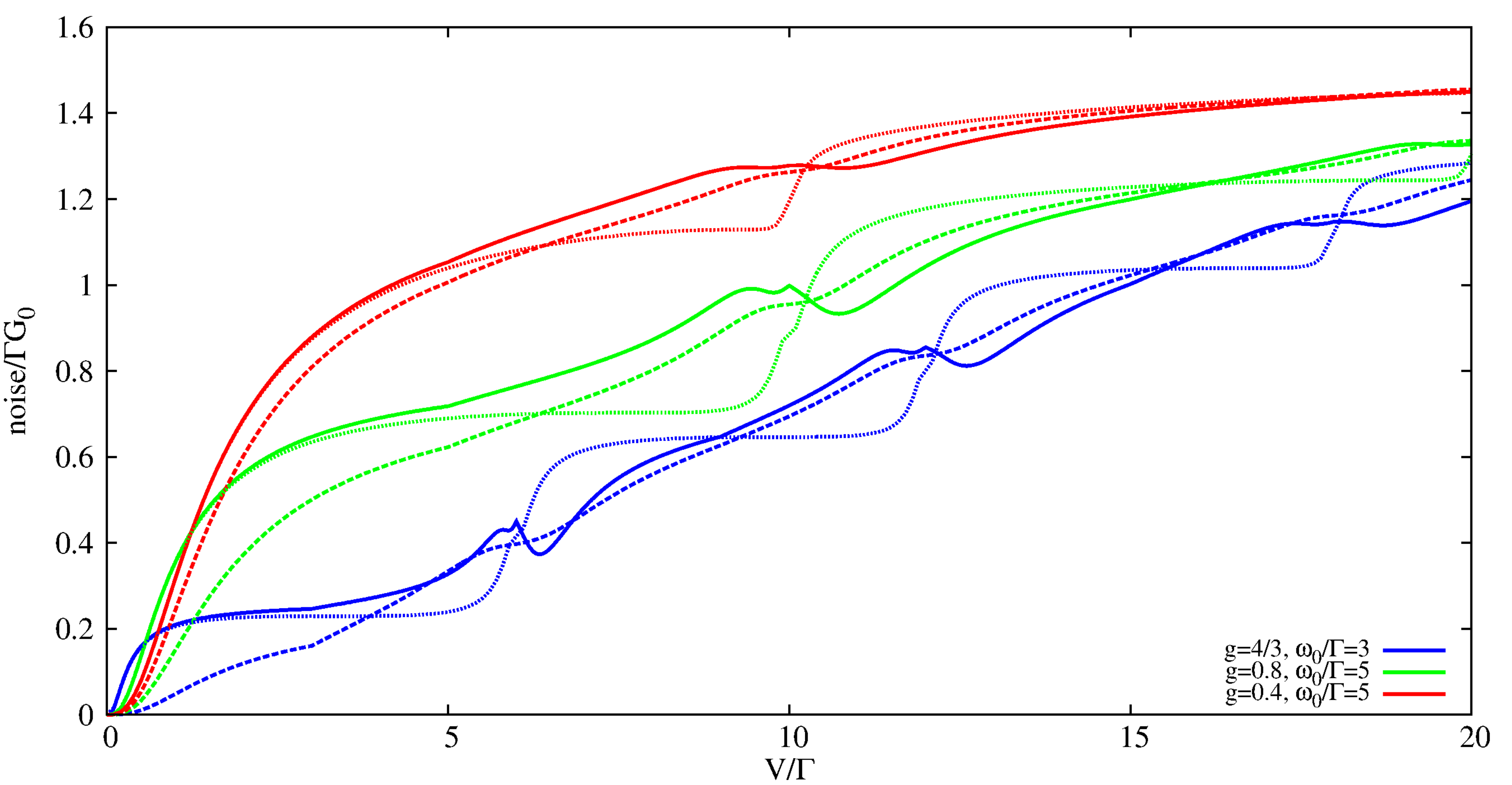}
  \caption{\small{Shot noise at $T=0$ for DTA (continuous line), SPA (discontinuous line) and PTA (dotted line) approximations for a symmetric case ($\tilde{\epsilon}=0$)}}
  \label{Tnoise}
\end{figure}

In this section we present the results computed from the general expressions of the shot noise for each approximation described in the section before. First comparison between the three approximations 
that can be done is at zero temperature. In this conditions the thermal noise is neglected, and it appears a noise depending only on the voltage applied to the junction. In figure \ref{Tnoise} some 
curves for the shot noise are presented. As it was expected from the behavior of the intensity, the noise at low bias voltage for the DTA and the PTA approximation are equal, while the SPA tends to 
describe a lower shot noise.\newline

Another important regime is the high voltage limit. In this limit we have probed that the three approximations tend to the same value of the shot noise, that depends only the value of the 
tunneling rate ($\Gamma$). This asymptotical value is the same that the one for the intensity at high voltage. This means that, in this conditions, the possible processes can be described through
a poissonian distribution.\newline

In this figure we can also observe how the PTA tends to describe stronger jumps in the shot noise, while in SPA these jumps appear just like a change in the slope of the curve. The DTA presents a 
different behavior. This approximation describes oscillations on the shot noise when the voltage reaches twice the value of the internal frequency.\newline

\begin{figure}
  \centering
	\includegraphics[width=1\textwidth]{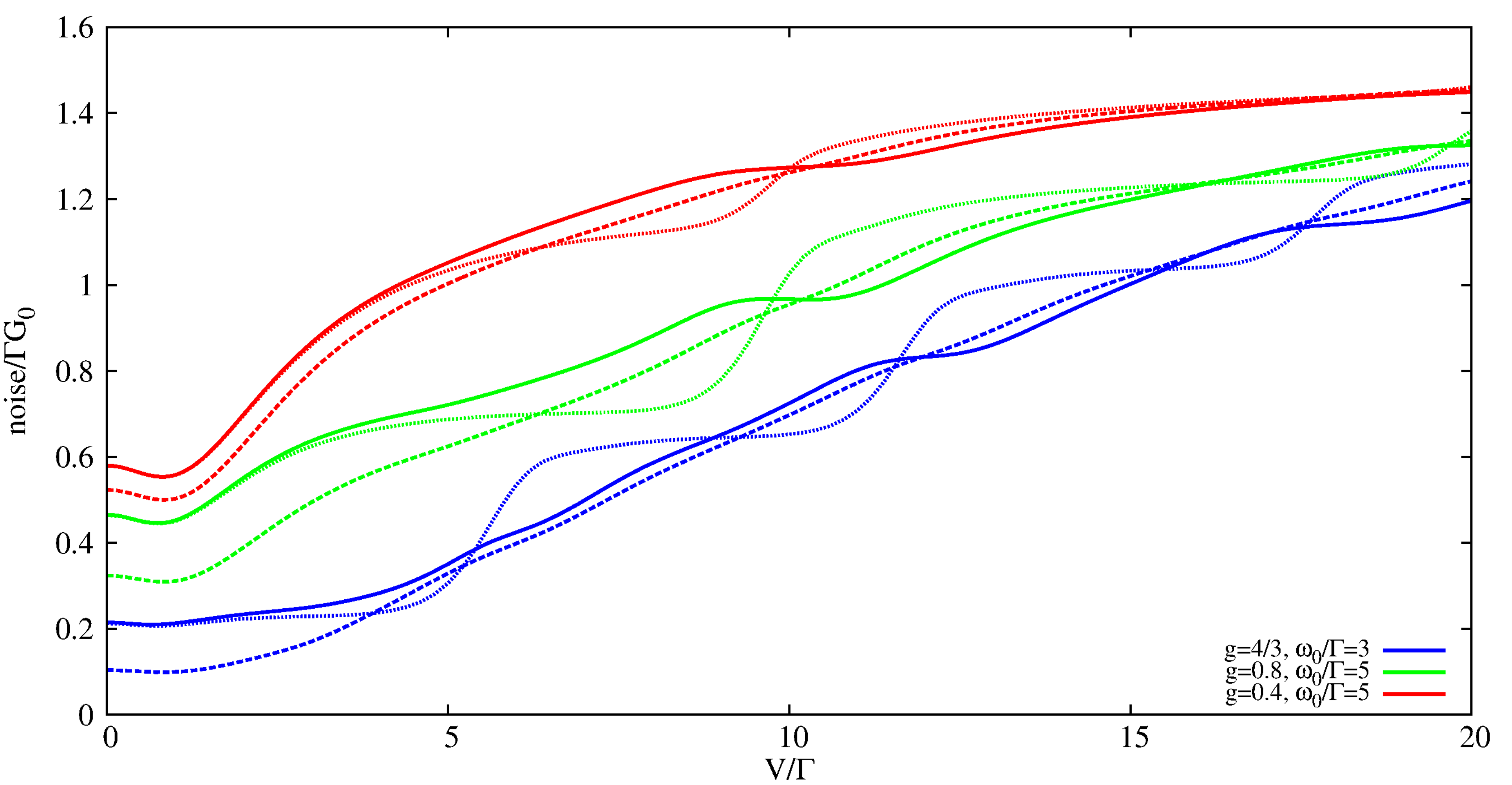}
  \caption{\small{Finite temperature shot noise for DTA (continuous line), SPA (discontinuous line) and PTA (dotted line) models and symmetric case ($\tilde{\epsilon}=0$) at $T=0.2/\Gamma$}}
  \label{fignoise}
\end{figure}

At finite temperature, there is a thermal component in the shot noise. It is due to possible thermal excitations and has already taken into account in the expressions of the shot noise described for each 
approximation. In figure \ref{fignoise} we present some noise curves at finite temperature.\newline

In comparison with the zero temperature curves, we can see that every approximation describes softer jumps. Moreover, the DTA loses its oscillations when reaching the frequency of the polaron. It
also shows an intermediate behavior between the PTA (that describes higher jumps) and the SPA (whose jumps are very small).
\newline

The value of the shot noise at zero bias voltage has changed, following the fluctuation-dissipation theorem. This theorem fixes the value of shot noise at zero voltage to be proportional to the 
conductance $S(V=0)=4TG(V=0)$. It has been checked for all three approximations. The SPA approximation starts from a lower value of the noise (because the lower conductance exhibited at zero bias), 
and it tends to approach to the rest of the curves at high voltage.\newline

Also at finite temperature, the three approximations preserve the same asymptotic limit. This means that at high voltage the noise is dominated by the effects of the voltage and poissonian statistics.

\section{Concluding remarks}
In this thesis a new approximation has been developed for studying the problem of electronic transport through a molecular junction with an internal vibrational degree of freedom for a strong
 polaronic regime. The proposed approach consists in a resummation of the dominant Feynman diagrams form the exact perturbative expansion. The results given for this approximation have
 been compared with other ones found in the literature in the polaronic and non polaronic regime. With this comparison we have concluded that this approximation is valid for both regimes and for a
 wide range of parameters.\newline

Also the shot noise has been studied comparing the approximations described in this work. Two known limits have been checked. The first one, at zero bias voltage, all the approximations follow the 
fluctuation-dissipation theorem. At high voltage the noise tends to be equal to the current intensity ($S=I$). This means that, in this limit, the statistics the charge transferred is possonian.\newline

There is much further work that can be done starting from this approximation. One interesting  problem that can be studied using our approximation is the time evolution of the system for a given initial
 condition \cite{Alfredo}. This calculation will let us to compare directly with numerical Monte Carlo simulations for time dependence.\newline

\section{Acknowledgments}
I would like to use this small space to to express my deepest appreciation to all those who provided me the possibility to complete this work. I wish to thank my parents, Jose Luis and Maria José,
 my sister Tania and my friends. Their love provided my inspiration and was my driving force.
I am in debt to K. F Albretch for sending me the quantum Monte Carlo simulations. I also would like to thank A. Zazunov for the very useful discussions.\newline

Last, but not least, I would like to thank my supervisors, Alfredo, Carmina and Álvaro, for the useful comments, remarks and engagement through the learning process of this master thesis.
Finally, I would also like to express my gratitude to \emph{Ministerio de Economía y Competitividad} for the financial support.

\end{document}